\documentclass{article}

\usepackage{arxiv}

\usepackage[utf8]{inputenc} % allow utf-8 input
\usepackage[T1]{fontenc}    % use 8-bit T1 fonts
\usepackage{hyperref}       % hyperlinks
\usepackage{url}            % simple URL typesetting
\usepackage{booktabs}       % professional-quality tables
\usepackage{amsfonts}       % blackboard math symbols
\usepackage{nicefrac}       % compact symbols for 1/2, etc.
\usepackage{microtype}      % microtypography
\usepackage{lipsum}		% Can be removed after putting your text content
\usepackage{graphicx}
\usepackage[numbers]{natbib}
\usepackage{doi}

\usepackage{xcolor}
\usepackage{graphicx}
\usepackage{amsthm}
\usepackage{mathtools}
\usepackage{amsmath, graphicx,multicol,multirow,array, bbm,varwidth,semantic}

\usepackage{subfig}
\usepackage{enumitem}

\newtheorem{theorem}{Theorem}[section]

\theoremstyle{definition}

\newtheorem{proposition}[theorem]{Proposition}

\newcommand{\din}{d^{in}}
\newcommand{\dout}{d^{out}}

\title{Fairness Rising from the Ranks: \\ HITS and PageRank on Homophilic Networks}

\date{} 					% Or removing it

\author{Ana-Andreea Stoica \\
	Max Planck Institute for Intelligent Systems,\\
	T{\"u}bingen, Germany
	\And
	Nelly Litvak \\
	Eindhoven University of Technology, \\
	Eindhoven, Netherlands \\
	\AND
	Augustin Chaintreau \\
	Columbia University, \\
	New York, USA \\
}

\renewcommand{\shorttitle}{Fairness Rising from the Ranks: HITS and PageRank on Homophilic Networks}

\hypersetup{
pdftitle={HITS and PageRank on Homophilic Networks},
pdfsubject={computer science},
pdfauthor={Ana-Andreea Stoica, Nelly Litvak, Augustin Chaintreau},
pdfkeywords={search algorithms, link analysis ranking, pagerank, hits},
}

\begin{document}
\maketitle

\begin{abstract}
  In this paper, we investigate the conditions under which link analysis algorithms prevent minority groups from reaching high ranking slots. We find that the most common link-based algorithms using centrality metrics, such as PageRank and HITS, can reproduce and even amplify bias against minority groups in networks. Yet, their behavior differs: one one hand, we empirically show that PageRank mirrors the degree distribution for most of the ranking positions and it can equalize representation of minorities among the top ranked nodes; on the other hand, we find that HITS amplifies pre-existing bias in homophilic networks through a novel theoretical analysis, supported by empirical results. We find the root cause of bias amplification in HITS to be the level of homophily present in the network, modeled through an evolving network model with two communities.  We illustrate our theoretical analysis on both synthetic and real datasets and we present directions for future work.~\looseness=-1
\end{abstract}

\section{Introduction}

Ranking algorithms govern the space of information retrieval, with a plethora of research in online search algorithms that can identify relevant and credible sources of information~\cite{page1999pagerank,kleinberg1999authoritative}. In the myriad of information sources permeating the web space, filtering through noise in the search for relevant data sources is a difficult task.~\looseness=-1

A recent line of work has pointed out the potential of such ranking algorithms to preference popularity over relevance~\cite{fortunato2006topical}. While users may genuinely be interested in popular items, popularity bias becomes an issue when it trades off with relevance. Moreover, popular items correlate to membership in an advantaged societal group: for example,~\citet{vlasceanu2022propagation} show that even gender neutral queries create disparate representation between men and women on Google search, as searching for the word `person' shows a disproportionate amount of men in the top results in Google images; in addition, the bias against women in search results correlates with the gender gap index of the country in which the experiment was run from. Another recent example shows that scientific mentorship networks have a vanishing number of women among those with high centrality, much lower than their general proportion~\cite{avin2015homophily}; thus, a ranking based on degree would amplify the underrepresentation of women in the field. Whether the bias was reproduced against demographic minority groups~\cite{espin2022inequality,vlasceanu2022propagation} or against newer agents that enter the system~\cite{mariani2015ranking,cui2022algorithmic}, empirical evidence amounts to a convincing argument that interaction data can amplify popularity among those who are already central in the network.  
Thus, ranking heuristics do not necessarily \textit{create} inequality between different demographic groups in the output ranking, but they may reproduce and even \textit{amplify} existing societal inequality.~\looseness=-1 

In this paper, we formally investigate structural causes for which link analysis ranking algorithms amplify, mirror, or reduce inequality between different social groups. We formally study the HITS algorithm~\cite{kleinberg1999authoritative} both theoretically and empirically, and compare its behavior to the PageRank algorithm. We employ a simple yet subtle evolving network model with multiple communities which encapsulates intrinsic inequality through preferential attachment dynamics and homophily. The model reproduces inequality between the degree centrality of different communities, with nodes from larger communities having larger degree, a trait often observed in real-world data. Such a generative model provides us with a tool for studying the true impact of algorithms on pre-existing inequality. As link analysis ranking algorithms have been developed as alternatives to the degree ranking, we use the degree distribution of different communities in our network as a benchmark, asking the question: \textit{if the degree ranking is a baseline for social capital inequality, when and why do link analysis algorithms induce an even more unequal distribution in the ranking scores, and when do they correct degree bias?}

The answer to these questions reveals subtle dynamics: different algorithmic choices and levels of inequality have a drastically different impact on the fairness of rankings outcome. We show that the interplay of high homophily and unequal degree distribution plays a central role in amplifying bias in rankings produced by HITS, through a new theoretical analysis based on the aforementioned model. We find that simulations on real-world data follow the predictions of the model. In comparison, a recent theory~\cite{antunes2023attribute} shows that PageRank corrects the degree bias for top ranked nodes. We confirm this theory with an extensive empirical analysis, finding that the bias correction does not translate beyond the top few most popular nodes. Instead, we find that the PageRank distribution closely follows the bias in the degree distribution in both synthetic and real data for the majority of nodes, illustrating limitations of recent theories. Our analysis paves the way for understanding differences in the impact of algorithmic choices on bias amplification. Indeed, we find that our results provide a first theoretical explanation: on the one hand, the difference between PageRank and HITS can be explained through a theory of random walks, showing that it is the reinforcing effect of backward-forward paths that HITS employs (as opposed to just forward walks in PageRank) that greatly amplifies bias in an already clustered network; on the other hand, random restarts in the random walks together with normalizing the influence of high indegree nodes provide a limited help in theory, as we show that it brings a ranking closer to the degree ranking (and therefore mirroring the degree bias); in practice, using randomization in real-world data improves fairness to a greater extent.

\paragraph{Our contributions:}

\begin{itemize}   
    \item We use a model of evolving networks that encodes bias in the degree distribution and find empirical evidence that PageRank reproduces the bias in the degree distribution, except for the top ranked nodes, for which it alleviates bias. (Section~\ref{sec:theory-hits})
    \item Inspired by this finding, we develop theory that predicts conditions for which bias is alleviated, reproduced, or amplified in link analysis algorithms, with a case study of HITS. We derive a closed-form condition showing that the level of bias amplification directly depends on how homophilic the network is---more clustered networks lead to a more biased ranking. To our knowledge, our results are the first to highlight differences between the impact on fairness of the PageRank and HITS algorithms. (Section~\ref{sec:theory-hits})~\looseness=-1
    \item We find evidence to support our findings in real-world data with varying levels of homophily. In practice, we find that PageRank can be less fair than predicted by theory. (Section~\ref{sec:bpamrankingexp})~\looseness=-1
    \item We characterize the role of randomization in improving fairness, providing empirical and theoretical evidence that randomization in HITS improves fairness. In essence, randomization leads to mirroring the degree bias or even improving upon it. We experiment with other variations of HITS based on multiple eigendirections, finding that fairness very much depends on the number of dimensions used and on the specific data. (Section~\ref{sec:hitsextensions})~\looseness=-1
\end{itemize}

These results open several avenues of research, as we find explanations for differences among classic algorithms that use network structures to provide a ranking of nodes. We conclude by proposing a series of future directions in how to progress in fair ranking using algorithmic design choices sensitive to the root cause of bias---whether it is a structural cause (in the form of centrality differences) or social (in the form of homophilic behavior).~\looseness=-1

\section{Background and Related work}
\label{sec:relwork}
\subsection{Background and modeling choices}

\paragraph{Link analysis algorithms.} Ranking algorithms that  leverage the connections formed between different information sources (websites, blogs, etc) were developed on the assumption that a link is equal to an endorsement. Popular sources gain their credibility through many links pointing towards them, weighted by different functions that capture how important downstream paths are. In this paper, we focus on two main algorithms, PageRank~\cite{page1999pagerank} and Hyperlink-Induced Topic Search (HITS)~\cite{kleinberg1999authoritative}, described below:~\looseness=-1

\textit{The PageRank algorithm} was developed to leverage random walks on graphs, with the intuition that the more these walks land on a node, the more important or central that node is~\cite{page1999pagerank}. Closely related to the use of degree centrality, the asymptotic behavior of these random walks governs their ranking position. Indeed, for undirected graphs, the stationary distribution exists and is proportional to the degree distribution. For directed graphs, the stationary distribution does not necessarily exist without a small modification, that of adding a random re-start, giving us the PageRank equation:  

\begin{equation}
    \textbf{x}_{t + 1} = \eta \cdot P \cdot \textbf{x}_t + (1 - \eta) \cdot \textbf{v},
    \label{eq:pagerank}
\end{equation}
where $P$ is the transition matrix, $1 - \eta$ is the restart probability, and $\textbf{v}$ is the vector of teleportation (taken to have all coordinates equal to $1/N$, where $N$ is the number of nodes in the network). For a graph $G(V,E)$ with a set of nodes $V$ and a set of edges $E$, denote its adjacency matrix by $A$ and note that $P_{ij} = \frac{A_{ij}}{\sum_i A_{ij}}$ ($\sum_i A_{ij}$ is the outdegree of a node $j$). Equation~\ref{eq:pagerank} is proven to have a stationary distribution $x_{t} \rightarrow x^{*}$ as $t \rightarrow \infty$, which is the coefficient vector denoting the importance of each node.~\looseness=-1

\textit{The HITS algorithm} was developed to account not only for those of high indegree~\cite{kleinberg1999authoritative}, but also to provide a sense of credibility by accounting for `hubs' (nodes of large outdegree). In doing so, what matters most in the HITS algorithm is whether one's neighbors are trusted sources of information. This trust is formalized through being a `hub' where information aggregates, contrasted with an `authority' to which many hubs point to. The algorithm formalizes this through a bipartite transformation of a graph, where each node $u \in V$ has now a hub score $h(u)$ and an authority score $a(u)$ that reinforce each other at every time step $t$ through the update equations:~\looseness=-1

\begin{equation}
\begin{gathered}
    a^{(t + 1)}(u) = \sum\limits_{v, (u,v) \in E} h^{(t)}(v) \mbox{ and } 
    h^{(t)}(u) = \sum\limits_{v, (u,v) \in E} a^{(t)}(v),
\end{gathered}
\label{eq:updateequations}
\end{equation}
where $a^{(0)}$ and $h^{(0)}$ are the initial conditions (often set to $1$, as in~\citet{kleinberg1999authoritative}). This set of equations is proved to converge to $a^{*}$ and $h^{*}$, respectively, which have the following properties: 

\begin{theorem}[Kleinberg, 1999]
$a^{*}$ is the principal eigenvector of $A^TA$ and $h^{*}$ is the principal eigenvector of $AA^T$.
\end{theorem}

In theoretically analyzing these algorithms, we employ a network model, called the Biased Preferential Attachment Model (BPAM), that reproduces commonly observed characteristics of biased networks: multiple communities, homophily, and a skewed degree distribution that induces a power-law distribution with \textit{different} exponents for different communities. In such a model, the degree ranking is inherently unequal: as we move towards higher ranks, the proportion of a minority group (with a lower exponent in the power law degree distribution) effectively vanishes, creating a so-called `glass ceiling effect' as measured through social capital~\cite{avin2015homophily}. We employ this terminology to investigate when a glass ceiling effect gets \textit{reduced} or \textit{amplified} in the ranking produced by link rank analysis algorithms, i.e. when does a minority get better or worse access to higher rankings. 
We find a continuous relationship between increased homophily and bias amplification 
in HITS in the BPAM through a mean-field analysis employing a theory of random walks.~\looseness=-1 

\subsection{Related work in fairness in ranking}

\paragraph{Fairness in link analysis ranking.} The question of fairness in link analysis algorithms has only recently started to gain attention.~\citet{tsioutsiouliklis2021fairness} study parameter regimes in which the total weight of PageRank for different communities satisfies statistical parity. In subsequent work,~\citet{tsioutsiouliklis2022link} propose a link recommendation algorithm aimed at improving the total weight of minority groups given by the PageRank algorithm. 
~\citet{espin2022inequality} analyze PageRank under network models and definitions similar to ours, defining fairness with respect to statistical parity and analyzing the distribution of ranks across different communities, i.e. when can a minority group accede a high rank under PageRank (without comparing to the degree distribution ranking).~\citet{antunes2023attribute} theoretically show that, in network models with homophily and a power-law degree distribution with different exponents among different groups, the PageRank score distribution for top nodes follows a power law with the same exponent for different communities; yet, this result does not capture the distribution of any other node that is not in tail of the degree distribution to begin with. In our work, we find new empirical evidence that PageRank only improves the ranking of minorities as compared to degree for the very top at best in synthetic data (Section~\ref{sec:theory-hits}), and may in fact entirely mirror the degree bias in real-world data (Section~\ref{sec:bpamrankingexp}). This result motivates us to understand the fundamental ways in which structural bias permeates link analysis algorithms.~\looseness=-1

On the other hand, HITS has received much less scrutiny with regard to general fairness questions, however with significant research done on its effectiveness (e.g.~\citet{najork2007hits} showing that it can outperform PageRank on the Web in terms of the NDCG score and the mean average precision). Previous empirical evidence has shown that completely disconnected components will have an odd effect on HITS, with larger groups being promoted~\cite{lempel2000stochastic,ng2001stable,Borodin:2005wo}---an effect known as the \textit{tightly knit community} (TKC) effect. Our results present a formal and more in-depth analysis of the TKC effect: we find that groups do \textit{not} have to be completely disconnected for minorities to suffer in the ranking; instead, there is a continuous link between homophily and minority underrepresentation in HITS. We explain this link through a theory of random walks, noting that high indegree nodes reinforce each other's authority scores. We show that renormalizing the influence of high indegree nodes and adding a random restart (equivalent to randomized HITS~\cite{ng2001stable} and conceptually similar to the SALSA algorithm~\cite{lempel2000stochastic}) alleviates the bias amplification, but only inasmuch as following the degree distribution. In previous work, randomized HITS~\cite{ng2001stable} and SALSA~\cite{lempel2000stochastic} show an empirical diversity-enhancing effect, but only for the top $5-10$ ranks. We investigate in the potential of such variations in improving fairness at all ranks---as opposed to just the first few---in Section~\ref{sec:hitsextensions}. To our knowledge, these results are novel and provide a first in-depth explanation of the relationship between inherent bias in a network and ranking algorithms. We conclude with an empirical analysis of other variations of HITS that use multiple dimensions in the eigenspace, motivated by stability results on the use of multiple eigenvectors~\cite{ng2001stable} and by recent work showing that one extra dimension can improve fairness in data representations through PCA~\cite{samadi2018price}. We find that, empirically, using multiple eigenvectors does not present a consistent path for fairness improvement, as it very much depends on number of eigenvectors and the data used. This analysis opens an avenue of research for finding a systematic pattern in which lower dimensions may improve fairness by better representing minorities in a higher-dimensional embedding.

\paragraph{Fairness in learning-to-rank and machine learning.} Fairness issues pertaining to the representation of minority groups in rankings become particularly important when centrality metrics (such as PageRank centrality and HITS authorities) are used as features in learning-to-rank algorithms~\cite{liu2009learning,dali2011learning,veloso2008learning,agarwal2006learning}. Drawing from a vast literature on fairness in supervised machine learning~\cite{angwin2016machine,Kleinberg:2017gt,dwork2012fairness,hardt2016equality,Feldman2015,zafar2017fairness,dong2023fairness}, fairness constraints in the form of statistical parity have been proposed in different learning-to-rank procedures~\cite{biega2018equity,celis2017ranking,beutel2019fairness,singh2019policy,zehlike2017fa,zehlike2020reducing}. Such constrains often come as a post- or in-processing technique, without modeling the generative process of feature distribution (for an extensive survey on fairness in learning-to-rank and related tasks, see~\cite{zehlike2022fairness} and~\cite{zehlike2022fairness2}). In this work, we tackle the way bias permeates the feature space, with a focus on link analysis ranking algorithms whose outcomes often get used as relevance features. On its own, the problem of fairness in link analysis is current and presents subtle challenges; beyond that, it has repercussions on the impact of feature bias on machine learning algorithms used in rankings. Thus, using evolving network models in link analysis ranking presents an opportunity for studying the role of homophily and degree distribution in bias amplification, absent in the learning-to-rank literature.~\looseness=-1

\section{A theory of bias in HITS}
\label{sec:theory-hits}

\subsection{Model description and preliminaries}

We first start by analyzing the PageRank and HITS algorithms on synthetic data generated from a model of network evolution that encodes pre-existing bias in the degree distribution, which we describe below. We follow with a theory predicting the conditions in which bias gets overly amplified in HITS, finding a closed-form relation involving homophily.

\paragraph{The Biased Preferential Attachment Model} This model is a variant of the Preferential Attachment Model that has been recently proposed~\cite{avin2015homophily}.
We define the model for the case of two communities, denoted by red and blue. Each node in the network is thus either blue (B) or red (R). These communities may represent demographic groups, interest groups, political affiliations etc. At each timestep, the network grows as:

\begin{itemize}
\item \textit{Minority-majority partition}: a new node $u$ enters the network and receives the label $R$ with probability $r$ and the label $B$ with probability $1 - r$. We assume that the red community is in minority, with $0 \leq r \leq 1/2$.

\item \textit{Preferential attachment (rich-get-richer)}: $u$ chooses a node uniformly at random and \emph{copies} one of its edges. This is equivalent to the new node connecting proportionally to the ending node's degree, $\mathbb{P}(v \mbox{ is chosen}) = d_t(v) / \sum\limits_{u \in V_t} d_t(u)$, where $d_t(x)$ denotes the degree of node $x$ at time $t$, and $V_t$ is the set of nodes in the graph at time $t$. Preferential attachment is what leads to a \textit{rich-get-richer} effect, where a few nodes are very well-connected and most nodes have very few connections---an effect observed in many online networks, such as the Web structure~\cite{barabasi1999emergence,barabasi2004network}.~\looseness=-1

\item \textit{Homophily:} if the new node has a different label than the node it chooses to connect to, the connection is accepted with probability $\rho$ and the process is repeated until an edge is formed. A value of $\rho$ closer to $0$ means a more homophilic (and therefore segregated) network, while a value of $\rho$ closer to $1$ means a more integrated network. The homophily parameter $0 \leq \rho \leq 1$ captures the tendency of people to connect more easily with those from the same community. Homophilic behavior has been observed in many real networks among different attributes (e.g., demographics, location, interest groups, professional collaborations)~\cite{McPherson:2001vg,bramoulle2012homophily}.

\end{itemize} 

Thus, as the network grows according to this model, exactly one node and one directed edge are added at each timestep. 
In this model, everyone has an outdegree of $1$ and a different indegree. We repeat this model $d$ times until everyone has an outdegree of $d$, for a choice of $d$. Just as in the Preferential Attachment Model, this model asymptotically leads to a power law distribution for each of the groups, with a different coefficient for each community~\cite{avin2015homophily}:

\begin{theorem}[Avin et al, 2015]
A graph sequence $G(n)$ generated through the Biased Preferential Attachment Model exhibits a power law degree distribution asymptotically: 

\begin{equation}
\begin{split}
    \mbox{top}_k(B) & \sim k^{-\beta(B)}, \\ 
    \mbox{top}_k(R) & \sim k^{-\beta(R)}, \\
\end{split}
\end{equation}
with $\beta(B) < 3 < \beta(R)$, where $\mbox{top}_k(R)$ and $\mbox{top}_k(B)$ denote the number of red and blue nodes with a degree of at least $k$, respectively.
\label{thm-avin}
\end{theorem}

These coefficients can be analytically computed from the model. When the network is completely integrated or completely segregated ($\rho \in \{0,1\}$), or the two subgroups are equal in proportion ($r = 0.5$), the coefficients of the power law for the degree distribution are also equal. For intermediate values of homophily and in the presence of a minority group $r \in (0,0.5)$, the coefficients will be different, leading to a so-called `glass ceiling effect'~\cite{avin2015homophily}: minority nodes have a vanishing fraction in the top degrees.~\looseness=-1

\subsection{Synthetic data analysis using BPAM}

\begin{figure*}
\centering
\subfloat[$r = 0.3, \rho = 0.1$]
{\includegraphics[width=0.3\textwidth] {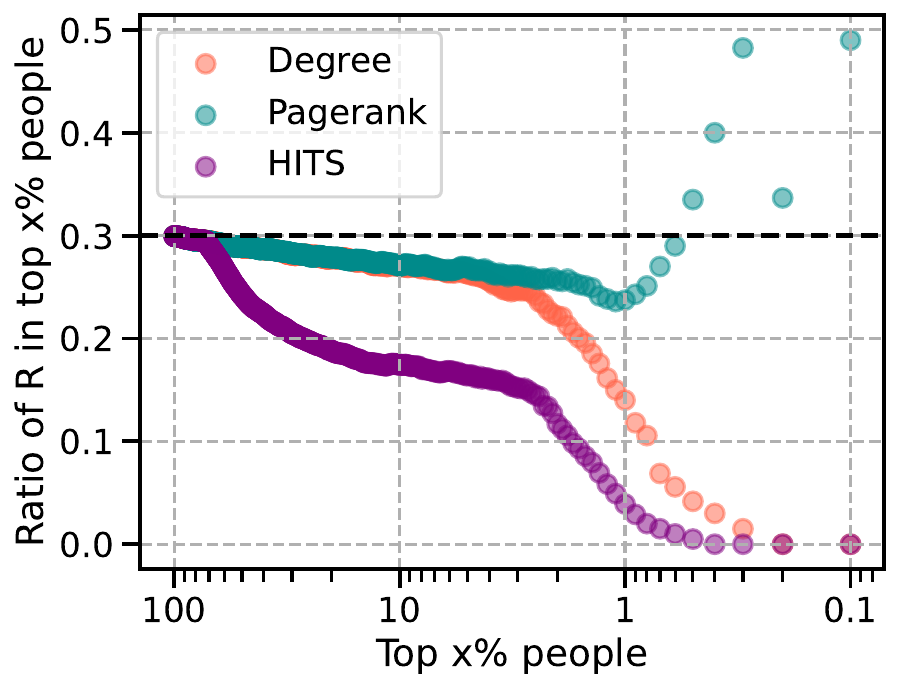}}
\subfloat[$r = 0.3, \rho = 0.3$]
{\includegraphics[width=0.3\textwidth] {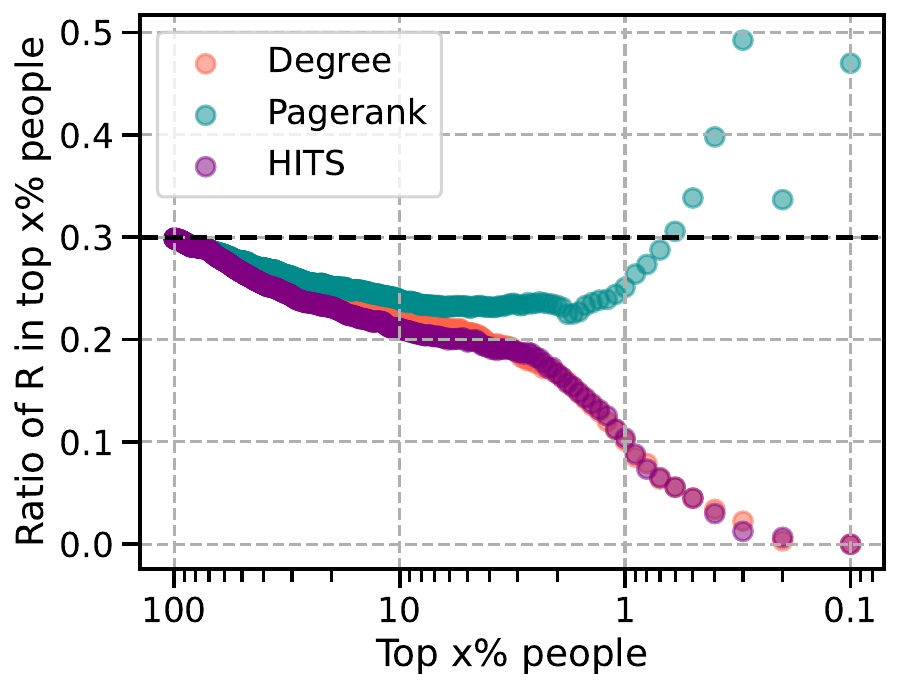}}
\subfloat[$r = 0.3, \rho = 0.5$]
{\includegraphics[width=0.3\textwidth] {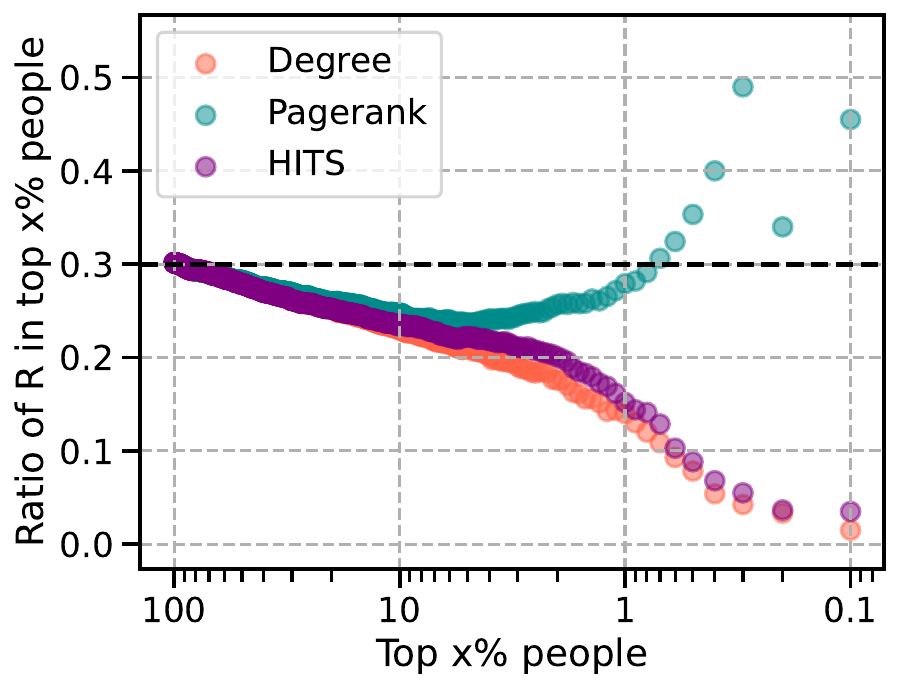}}
\caption{Representation of the minority group R in the ranking of the nodes based on degree (orange), HITS (purple), and PageRank (blue), for directed networks simulated from the BPAM with $N=1,000$ nodes and outdegree $d = 6$.}
\label{fig:ranking_symhomophily_directed}
\end{figure*}

We simulate the BPAM for $N=1,000$ nodes and outdegree $d=6$. For each network instance simulated from this model, we compute the degree distribution, the PageRank, and the HITS authority scores and we plot the ratio of minority members among those in the top $x$ percentange of ranking and above on the $x$-axis (plotted in log-scale), and adding a dashed horizontal line for the population minority ratio. All simulations are averaged over $1,000$ iterations. For example, the leftmost plotted point in Figure~\ref{fig:ranking_symhomophily_directed} is the ratio of minority members among the entire population, top $100\%$ (so, top $1,000$ people). In a sense, the dashed line is our fair baseline, also equivalent to achieving statistical parity: the further away the minority ratio plotted is from the dashed (often smaller), the less fair the ranking becomes. We know from Theorem~\ref{thm-avin} that the proportion of minority members among top ranked people by degree goes to $0$, an illustration of what a glass ceiling effect may look like in practice. We note that for $r = 0.5$ or $\rho = 1$, the degree distribution is the same among the two communities, and then the PageRank and HITS distributions are also fair (as in, are equal to or very close to the proportion of minority nodes in the population represented by the black dashed line). 
The closest $\rho$ gets to $0$ (Figure~\ref{fig:ranking_symhomophily_directed} a), the two communities are getting increasingly disconnected, and we note that HITS becomes progressively more unfair, while degree and PageRank become more fair. For example, while we have $30\%$ of minority proportion in the population, there are under $20\%$ among those in top $10\%$ and all above ranks in HITS. For moderate homophily (e.g. $\rho$ equal to $0.3$ and $0.5$ in panels b and c), HITS and degree ranking are quite similar, both leading to a vanishing minority fraction among the top ranked. For PageRank, while finding novel evidence that it is similar to the degree ranking for most of the ranking, we also find improvement in the minority ratio for the very top ranked nodes (note that~\citet{antunes2023attribute} proves that PageRank is fair only for the very top nodes). This shows that PageRank and HITS do not generally introduce bias when the degree is unbiased, but rather reproduce the bias that is already existing. Yet, they differ in how much of that bias they reproduce, with HITS being particularly sensitive to the level of homophily in the network.~\looseness=-1

\subsection{Bias in HITS: a Mean-Field Analysis}

We complement our experimental results with a theoretical analysis of the HITS algorithm to elucidate the structural reasons behind the amplification of bias in the rankings that HITS produces. Specifically, the BPAM helps accurately expound the role of homophily. Several works argue that the TKC effect reduces the diversity of search results in the top $5-10$ queries, and conjecture that this effect emerges when communities are completely disconnected in the larger network~\cite{lempel2000stochastic,ng2001stable,Borodin:2005wo}. We find a more subtle effect: even in connected networks, HITS can amplify the underrepresentation of minorities in the ranking outputted as compared to the degree ranking, and moreover, the more homophilic the network is, the more pronounced the bias against minorities.~\looseness=-1

\begin{theorem}
For a network $G(V,E)$ drawn from BPAM with $N >> 1$ nodes, two communities, minority ratio $r$, and homophily parameter $\rho$, the following hold: 

\begin{enumerate}
    \item the following inequalities are true under a mean-field approximation for the authority scores $a(\cdot)$ obtained from the HITS algorithm:
\begin{equation}
\begin{gathered}
    \overline{a(R)} <  \overline{a(B)} \mbox{ and } \overline{a(R,k)} < \overline{a(B, k)},
\end{gathered}
\label{eq:hitsthm}
\end{equation}
for $r \in (0, 0.5) \mbox{ and } \rho \in (0, 1)$, where $\overline{a(C)}$ (and $\overline{a(C,k)}$, respectively) denotes the average authority score of community of color $C$ (and of nodes of indegree equal to $k$, respectively).

\item for nodes $u \in R$, $v\in B$ of indegree equal to $k$, the ratio $\frac{a^{(t)}(u \in R)}{a^{(t)}(v \in B)}$ is an increasing function in the homophily parameter $\rho$ for any $t \geq 2$ and $k \geq 1$, under a mean-field approximation. 
\end{enumerate}
\label{thm:mainthm}
\end{theorem}
%The proof of this theorem relies on the properties of the model as well as on being able to interpret HITS in terms of backward-forward paths on graphs through a mean-field analysis. 
Equality in part one of Theorem~\ref{thm:mainthm} holds for $r \in \{0,0.5\}$ or $\rho \in \{0,1\}$. We provide a sketch of the proof in this section. The crux of the proof is in using the model to compute an approximation for the the authority score of nodes as a product of the nodes' indegree and an additional factor, termed here as a multiplicative factor. By analyzing the behavior of the multiplicative factor as a function of the model parameters, we can investigate how the authority score distribution compares to the degree distribution. First, we note that the authority score of the nodes after $t$ iterations of the HITS algorithm is proportional to the number of paths that alternate between backward and forward direction of the edges, starting from each node~\cite{Borodin:2005wo} (initializing with hub scores equal to $1$, then performing a first update on the authority scores as per equation~\eqref{eq:updateequations}). We can then approximate the number of backward-forward paths for nodes belonging to each community using the properties of the model. We detail our approximation in the extended version of the paper, noting that it boils down to identifying a dominant term in probability in a recurrent equation over the iterations $t$. We denote the approximation of the HITS at iteration $t$ by:~\looseness=-1

\begin{equation}
\begin{gathered}
    a^{(t)}(v \in R) \approx \din(v) \cdot (d-1) \cdot MF^{(t)}(R),\\
    a^{(t)}(v \in B) \approx \din(v) \cdot (d-1) \cdot MF^{(t)}(B),
\end{gathered}
\label{eq:recurrence-hits-factor-og}
\end{equation}
where $\din(v)$ is the (random) indegree of a vertex $v$, $d$ is the constant outdegree of a vertex, and $MF^{(t)}(R)$ and $MF^{(t)}(B)$ are the (random) multiplicative factors that allow to compare the HITS ranking with the degree ranking. We note that $\din(v)$ is $O_P(1)$ in $N$, since the fraction of vertices with degree of higher order of magnitude is vanishing, where $O_P(\cdot)$ means that the big-O relation holds in probability. To prove the first part of the theorem, we show the following results:~\looseness=-1 

\begin{proposition}
    For any $t \geq 2$, $MF^{(t)}(B) \geq MF^{(t)}(R)$, for all $0 \leq r \leq 0.5$ and $0 \leq \rho \leq 1$.
    \label{prop:mf-ineq}
\end{proposition} 

Then, knowing that the average indegree of a red node is lower or equal to the average indegree of a blue node (Theorem $4.1$, part $1$ in~\citet{avin2015homophily}), we get that by averaging over the set of red and blue nodes, respectively, $\overline{a(R)} \leq \overline{a(B)}$. Similarly, averaging over degree classes (all red nodes and all blue nodes of indegree equal to $k$, respectively, for all $k$), we get that $\overline{a(R,k)} \leq \overline{a(B, k)}$, which shows the first part of the theorem.

We note that the multiplicative factors are in fact functions of all the model parameters, including $r$, which we consider fixed. As noted in the extended proof, the bias against the minority group amplifies as $r$ gets closer to $0$. To prove the second part of the theorem, we consider the multiplicative factors as functions of the homophily parameter $\rho$ and define $F^{(t)}(\rho) := \frac{MF^{(t)}(R,\rho,r)}{MF^{(t)}(B,\rho,r)}$ (as $r$ is fixed). We note that for nodes with the same indegree, the ratio of their authority scores boils down to the ratio of the multiplicative factors in the approximation. Thus, we show the following:~\looseness=-1

\begin{proposition} 
    Define $F^{(t)}(x):= \frac{MF^{(t)}(R,x,r)}{MF^{(t)}(B,x,r)}$ as a function of the homophily parameter $\rho \in (0,1]$, for $r$ fixed. Then, 
    $F^{(t)}(x)$ is an increasing function in $(0,1]$ with $F^{(t)}(1) = 1$ for any $t \geq 1$.
    \label{prop:homophily_worse_hits_n}
\end{proposition}

\paragraph{Intuition:} The intuition that stems from this analysis is as follows: $F^{(t)}$ computes the ratio of the multiplicative factor of red and blue nodes in determining their authority score. When $F^{(t)}(x) = 1$, it implies that both communities have the same multiplicative factor, rendering the HITS ranking the same as the degree ranking. That means that nodes of similar degree will have a similar chance of showing up in a ranking of $k$ nodes regardless of color, and thus, by aggregation, the ratio of R nodes in the top k in the HITS ranking will be the same as in the degree ranking. When $F^{(t)} \neq 1$, that means that one community is `bumped up' (or `bumped down') more than the other, which will change the ranking. For example, if the red community gets a discount factor ($MF^{(t)}(R) < 1$) and the blue community gets a boosting factor ($MF^{(t)}(B) > 1$), that means that the red community is `pushed down' in the ranking as more blue nodes are overtaking red nodes that initially had similar or even better degree in the HITS ranking. If both communities get either a discount or a boosting factor, whichever factor is higher will `boost' that community upper in the ranking as compared to the original degree ranking. Thus, if the red community consistently has a lower multiplication factor than the blue community, that means that they are consistently `pushed down' in the HITS ranking as compared to the degree distribution. Proposition~\ref{prop:homophily_worse_hits_n} essentially shows that more homophilic networks experience exacerbated bias against minority (red) nodes compared to the initial degree distribution, while integrated communities will mirror the bias of the degree distribution.~\looseness=-1

The fundamental difference between HITS and PageRank lies in the way nodes aggregate centrality: while for PageRank, the transition matrix is equivalent to taking random walks in the `forward' direction starting from a node (following its outdegree), with a random restart, for HITS, the update equations are equivalent to taking `backward-forward' paths starting from a node, without any restart. The outdegree is constant in our model and the random restart in PageRank serves in discounting paths with exponential decay in the restart probability (for a full theoretical analysis of PageRank, see~\citet{antunes2023attribute}). On the other hand, the backward-forward paths mean that nodes with high indegree are greatly advantaged, amplifying their authority from other high indegree nodes with whom they have a commoon neighbor that points to both (in branching processes terminology, a common `child'). These paths are not discounted and serve to amplify the scores of already well-connected nodes, who are likely to connect with others of the same color due to homophily.

%%%%%%%%%%%%%%%%%%% SYNTHETIC DATA %%%%%%%%%%%%%%%%%%%
\section{Experimental evidence of bias}
\label{sec:bpamrankingexp}

%%%%%%%%%%%%%%%%%%% REAL NETWORKS %%%%%%%%%%%%%%%%%%%
% \subsection{Real data experiments}
\label{sec:realdataranking}

We analyzed several network datasets that contain directed edges, varying levels of homophily, and two main communities: a majority and a minority. In each of these datasets, either the majority or the minority community has the `advantage' in the degree distribution (meaning that they are over-represented at the top of the degree distribution, as the glass ceiling definition formalized in~\citet{avin2015homophily}). Although BPAM only models minorities with a degree \textit{disadvantage}, in reality either group can have a degree advantage, especially when the minority size is not far from $50\%$. The datasets are described below, color-coded by which group has the degree advantage (blue for majority, red for minority).\footnote{Real and synthetic datasets and the code used for this paper can be found at: \url{https://github.com/astoica/fairness_link_analysis_ranking}}~\looseness=-1

\begin{enumerate} 
    \item {\color{blue}\textbf{APS:}} The APS citation network~\cite{lee2019homophily} contains $1,281$ nodes, representing papers written in two main topics: Classical Statistical Mechanics (CSM), constituting $31.7\%$ of the papers, and Quantum Statistical Mechanics (QSM), accounting for the rest of $68.3\%$ of the papers. As~\citet{lee2019homophily} analyze, the dataset has high homophily, meaning that each subfield cites more papers in their own field than in the other field.~\looseness=-1
    \item {\color{blue}\textbf{DBLP:}} The DBLP citation dataset of computer scientists was collected from the DBLP platform~\cite{ley2009dblp,Tang:08KDD,sinha2015overview}, an online database that records most publications in computer science. We use version V10 of the dataset, using the authors present in all papers as nodes and associating a perceived gender of each node through the genderize.io API; we retain the nodes for which the probability of a gender is over $90\%$. A directed edge is created between two authors if a paper by the first author cites a paper by the second author. We extract the largest weakly connected component, obtaining a graph with $1,224,996$ nodes and two communities, men ($88\%$) and women ($22\%$). The data has low homophily.  
    \item {\color{red}\textbf{Instagram:}} An interaction network from Instagram collected by~\citet{stoica2018algorithmic} containing $553,628$ nodes and $652,931$ edges, where everyone has a labeled gender ($45.57\%$ men and $54.43\%$ women). Each edge between two users represents a `like' or `comment' that one user gave another on a posted photo. The data has moderate homophily.~\looseness=-1
\end{enumerate}

\begin{table}
\centering
  \caption{Data characteristics.}
 \label{table:data-ranking}
	\begin{tabular}{cccccccl}
    \toprule
      & Nodes & \% minority
      & Edges & HRI \\
    \midrule
    {\color{blue}APS} & 1,281 & 31.7 \%  & 3,064 &0.12\\
    {\color{blue}DBLP} & 1,224,996 & 22 \% & 95,160,219 & 0.74\\
    {\color{red}Instagram} & 539,023 & 45.6 \% & 640,211 & 0.44 \\
  \bottomrule
\end{tabular}
\end{table}

\begin{figure*}
\centering
\subfloat[APS]
{\includegraphics[width=0.3\textwidth] {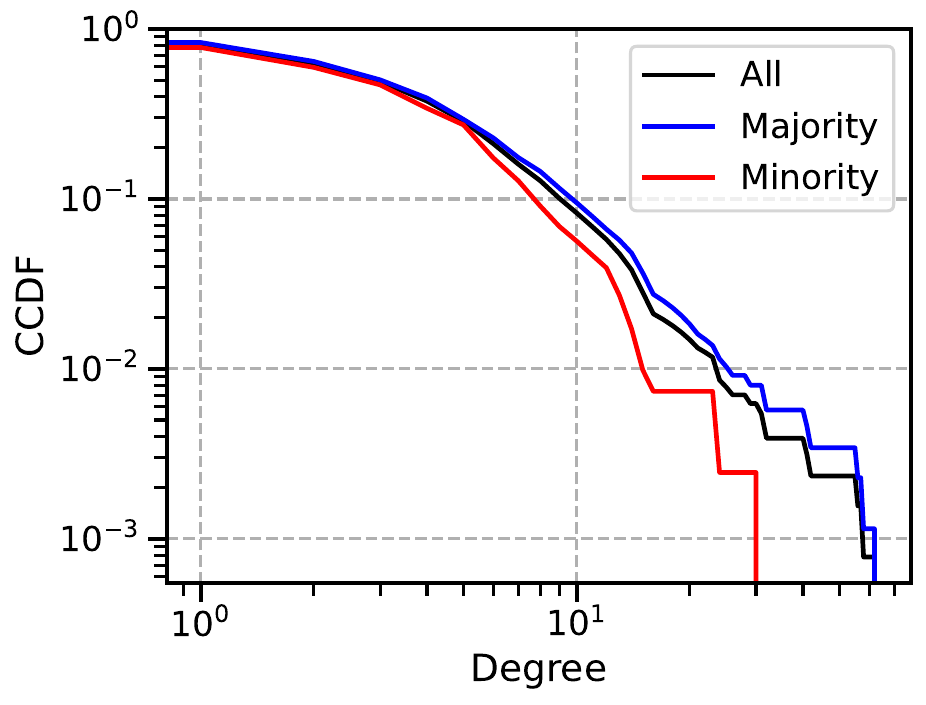}}
\subfloat[DBLP]
{\includegraphics[width=0.3\textwidth] {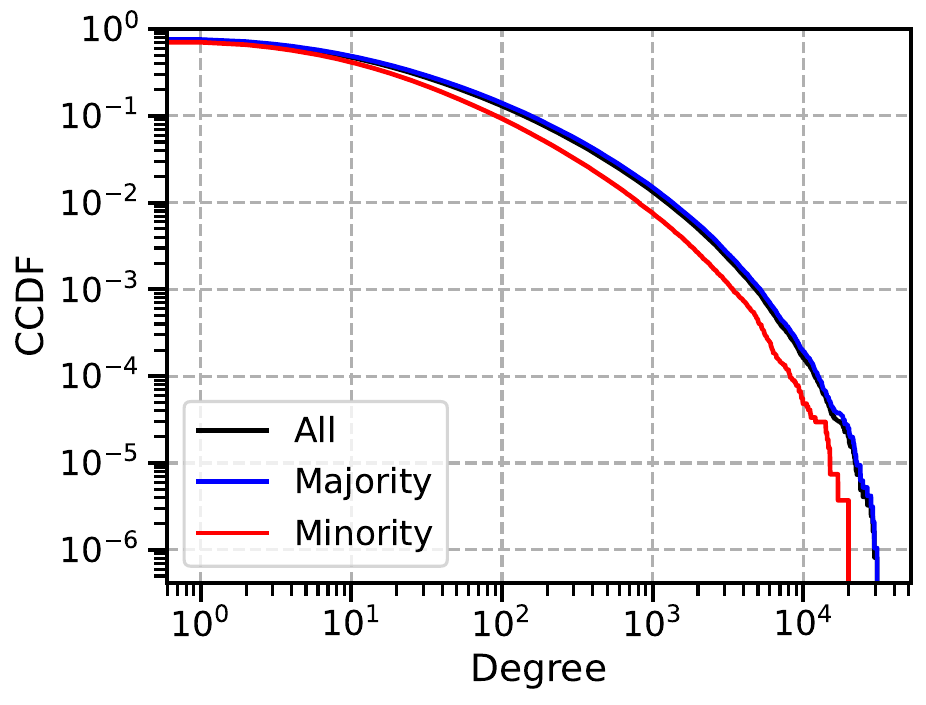}} 
\subfloat[Instagram]
{\includegraphics[width=0.3\textwidth] {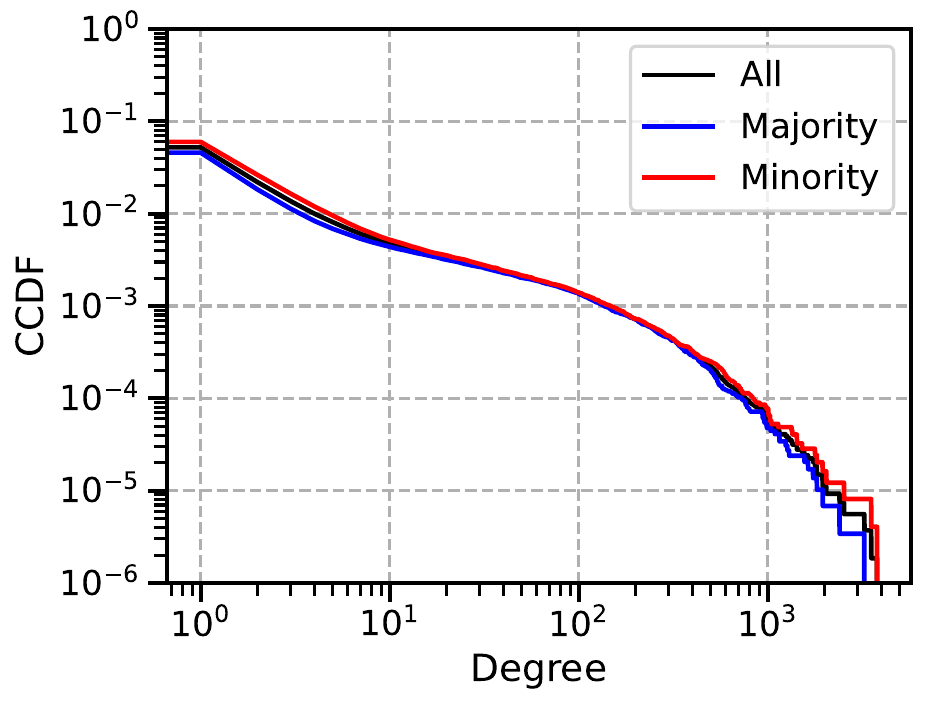}}
\caption{CCDF of the degree distribution of the minority group (red) and the majority group (blue) for the real-world networks. The black line illustrates the CCDF for the total population.}
\label{fig:ccdf_degree_realdata_ranking}
\end{figure*}

Table~\ref{table:data-ranking} presents data characteristics, including the number of nodes, edges, minority percentage, and an estimated homophily factor. Each network either has an advantaged majority group or an advantaged minority group, illustrated in Figure~\ref{fig:ccdf_degree_realdata_ranking}: when the complementary cumulative distribution function (CCDF), plotted in log-log scale, of a group is higher than the other on an interval, then the representation of that group for the degree classes belonging to that interval is higher than for the other group. When that interval includes the top of the degree hierarchy, it is used as evidence to illustrate a glass ceiling effect against one group.~\looseness=-1

In order to assess whether a network is homophilic or not, we use the \textbf{homophily rarefaction index} (HRI)~\cite{zhang2021chasm}, which computes the fraction of cross-community edges over the expected number of cross-community edges that the BPAM gives, which is $2 r (1-r) |E|$, where $E$ is the set of edges in the network. Other methods are the Newman assortativity index~\cite{newman2003mixing} and the asymmetric homophily index~\cite{lee2019homophily}. An HRI closer to $1$ means a more integrated dataset, while a lower HRI indicates higher homophily. The APS and DBLP datasets both have a majority group that has an advantage in the degree distribution, with APS being the most homophilic and DBLP the least. The Instagram dataset has a minority group that has a degree advantage and moderate homophily.~\looseness=-1

\begin{figure*}
\centering
\subfloat[APS]
{\includegraphics[width=0.3\textwidth] {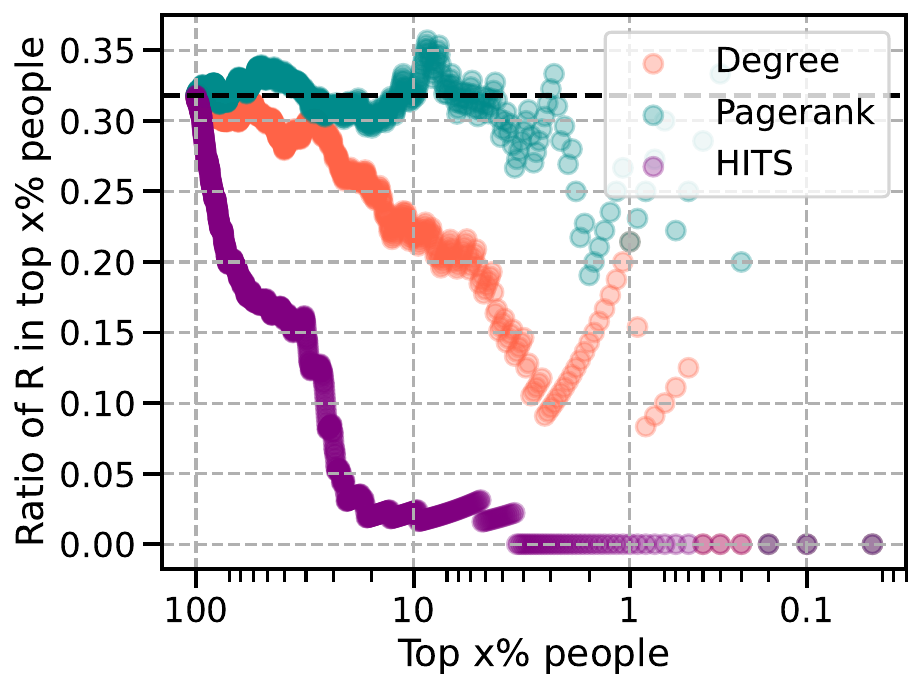}}
\subfloat[DBLP]
{\includegraphics[width=0.3\textwidth] {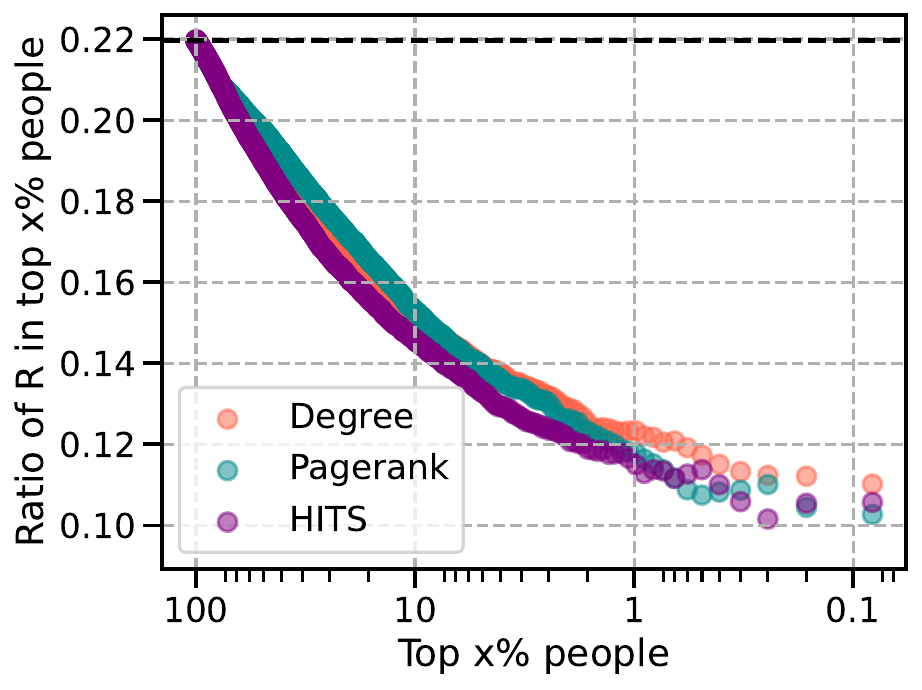}} 
\subfloat[Instagram]
{\includegraphics[width=0.3\textwidth] {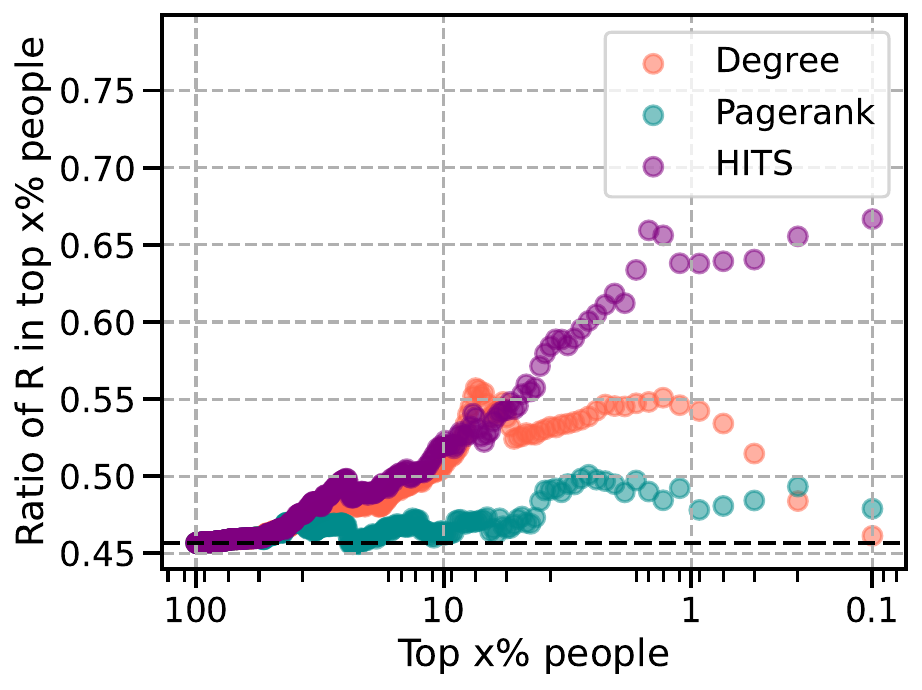}}
\caption{Representation of the minority group R in the ranking of the nodes based on degree (orange), HITS (purple), and PageRank (blue), for the real-world networks.}
\label{fig:ranking_realdata}
\end{figure*}

We tested PageRank and HITS on all three datasets, illustrating the ratio of the minority group among each rank in Figure~\ref{fig:ranking_realdata}. We notice that the data captures the dynamics predicted by the BPAM: as APS and DBLP have an advantaged majority group, the minority group gets underrepresented in the degree, and even more so in HITS. For APS, HITS (purple) amplifies the under-representation of the minority the most as compared to the degree distribution (orange), which is consistent with APS being the most homophilic network. DBLP is the least homophilic, consistent with HITS slightly amplifying the degree ranking bias. Instagram, on the other hand, has an advantaged minority group by degree, which is also over-represented in the ranking produced by both PageRank and HITS for most ranks. Instagram is more homophilic than DBLP but less than APS, explaining why HITS amplifies more of the minority advantage in Instagram than it amplifies the majority advantage in DBLP, and less than in APS. Although Therefore, HITS amplifies the advantage of whichever group has the degree advantage, while PageRank reduces this advantage: the majority in APS and DBLP, and the minority in Instagram.~\looseness=-1

These experiments show that the BPAM is relatively accurate in reproducing the data behavior on ranking algorithms. Furthermore, our experiments reveal the subtle effect of homophily, showing that in more homophilic networks, HITS amplifies bias against minorities. We note that while PageRank is provably fair for top degree nodes~\cite{antunes2023attribute}, it actually reproduces the degree distribution bias on DBLP, while preserving statistical parity for APS and Instagram for most of the ranking (except the very top, where it inherits some bias). This points to a cause of inequality stemming from the \textit{outdegrees} this time (as they are the normalizing factor of PageRank, and are not constant in the real data, like in the synthetic data generated from the BPAM), prompting further future investigations.~\looseness=-1

\section{Randomization: a path to fairness?}
\label{sec:hitsextensions}

Our findings so far motivate the search for algorithmic variations that can improve fairness. A natural question arises: can we de-bias HITS by normalizing the influence coming from the indegrees, and would a restart in the random walk (similar to PageRank) additionally help? Such a variation, named randomized HITS~\cite{ng2001stable}, has been shown to have perform well in terms of query relevance, yet has not been analyzed with respect to fairness for different communities.~\looseness=-1

\begin{figure*}[!ht]
\centering
\subfloat[$r = 0.3, \rho = 0.1$]
{\includegraphics[width=0.3\textwidth] {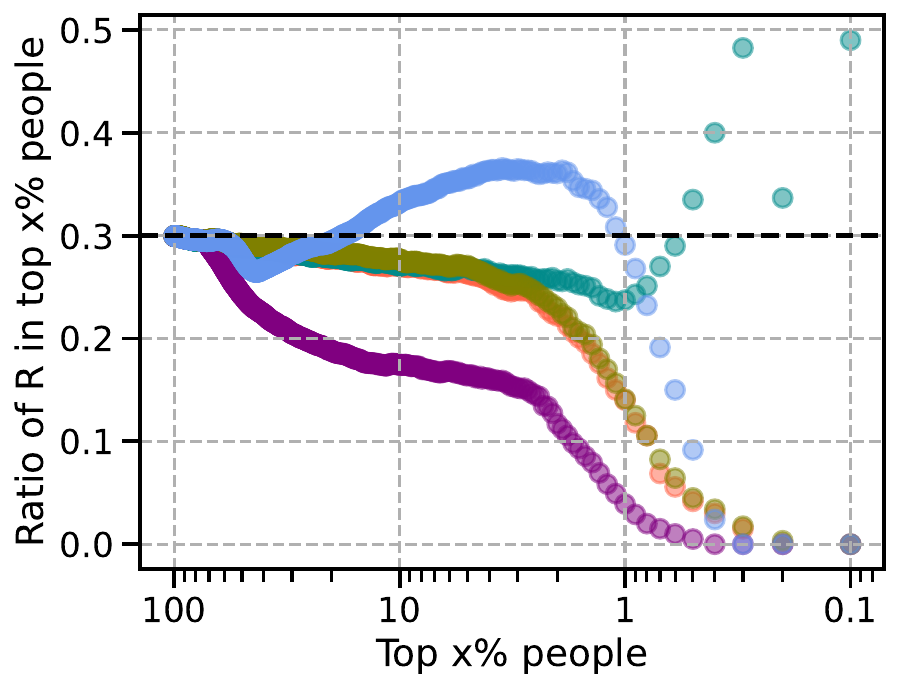}}
\subfloat[$r = 0.3, \rho = 0.3$]
{\includegraphics[width=0.3\textwidth] {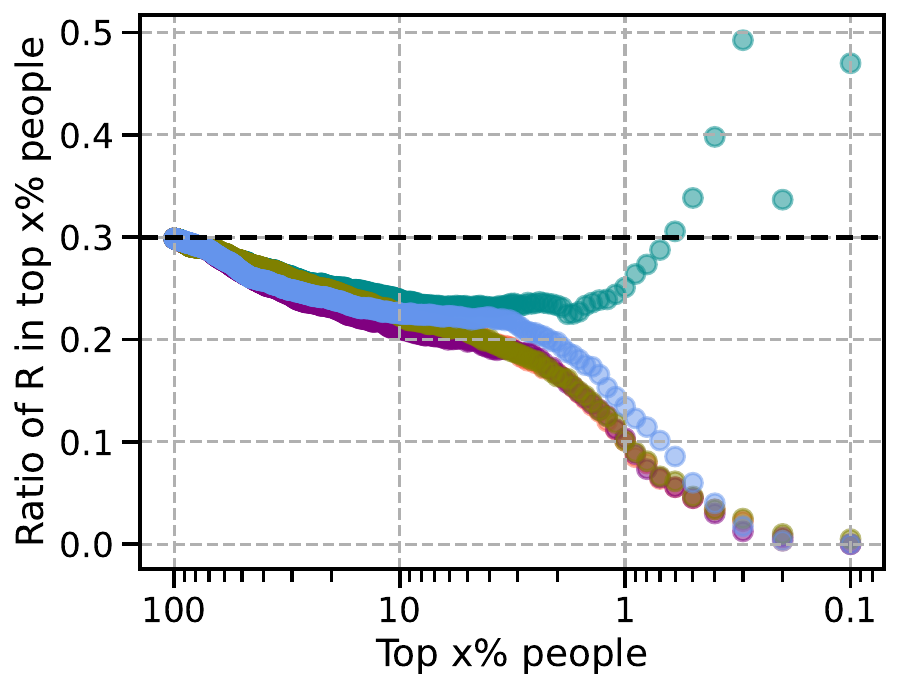}}
\subfloat[$r = 0.3, \rho = 0.5$]
{\includegraphics[width=0.3\textwidth] {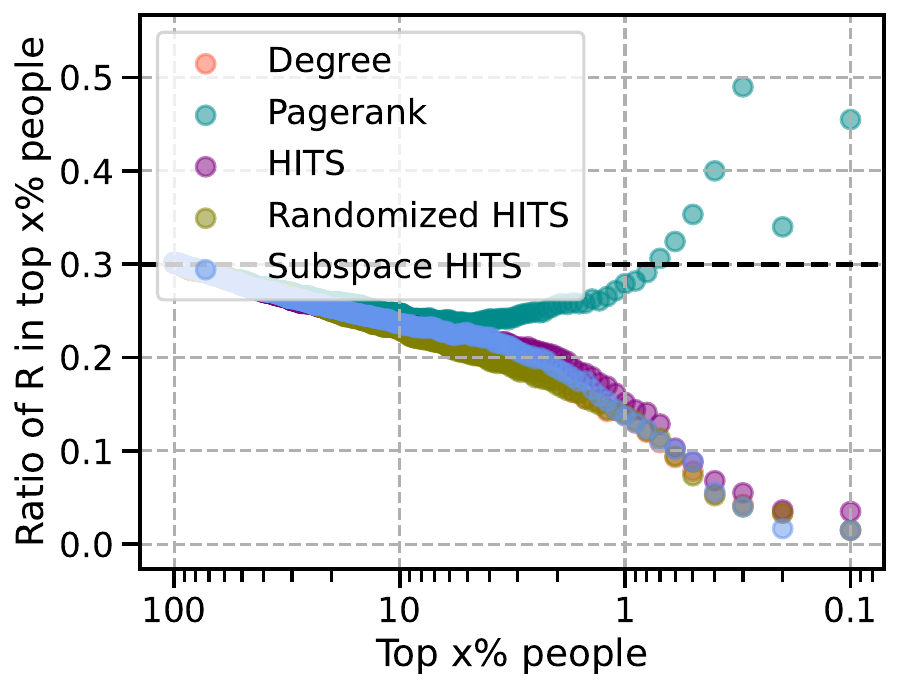}}
\caption{Representation of the minority group R in the ranking of the nodes based on degree (orange), HITS (purple), randomized HITS (olive), and PageRank (blue), for directed networks simulated from the BPAM with $N=1,000$ nodes and outdegree $d = 6$.}
\label{fig:randomHITS_ranking_symhomophily_directed}
\end{figure*}

\begin{figure*}[!ht]
\centering
\subfloat[APS, $f(\lambda) = 1$]
{\includegraphics[width=0.3\textwidth] {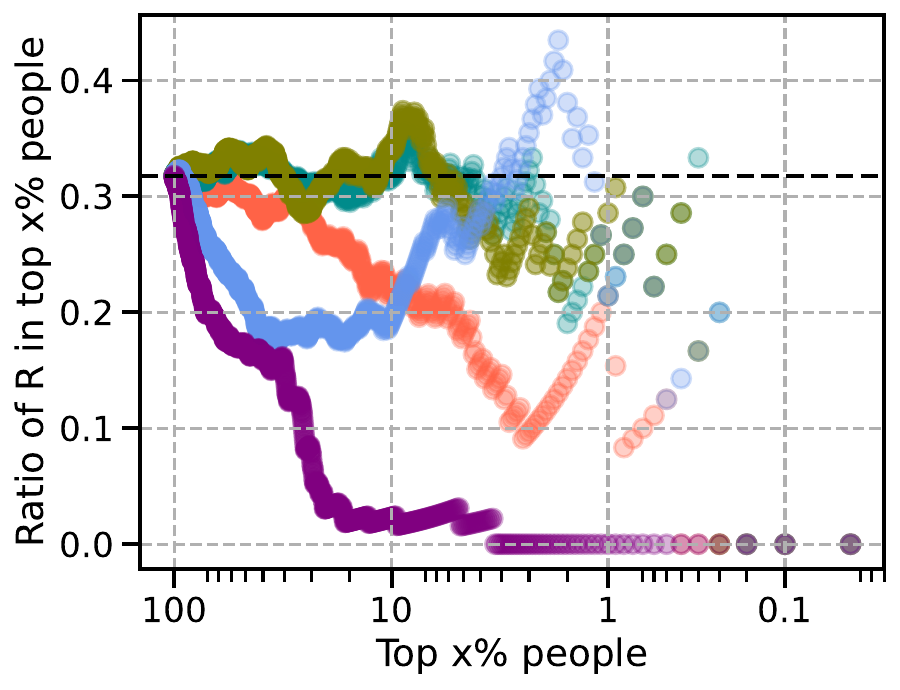}}
\subfloat[DBLP, $f(\lambda) = 1$]
{\includegraphics[width=0.3\textwidth] {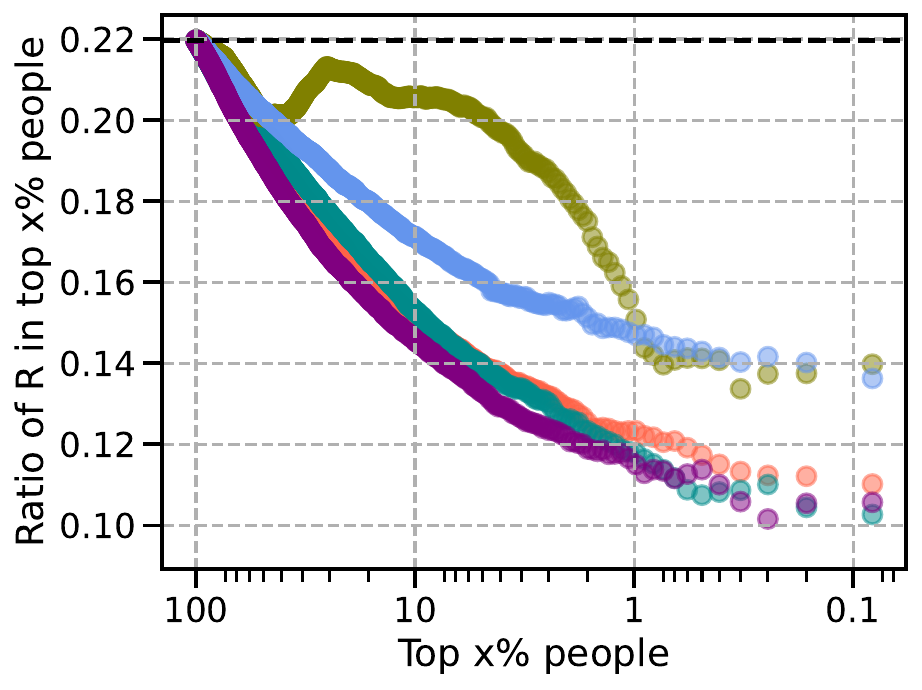}}
\subfloat[Instagram, $f(\lambda) = 1$]
{\includegraphics[width=0.3\textwidth] {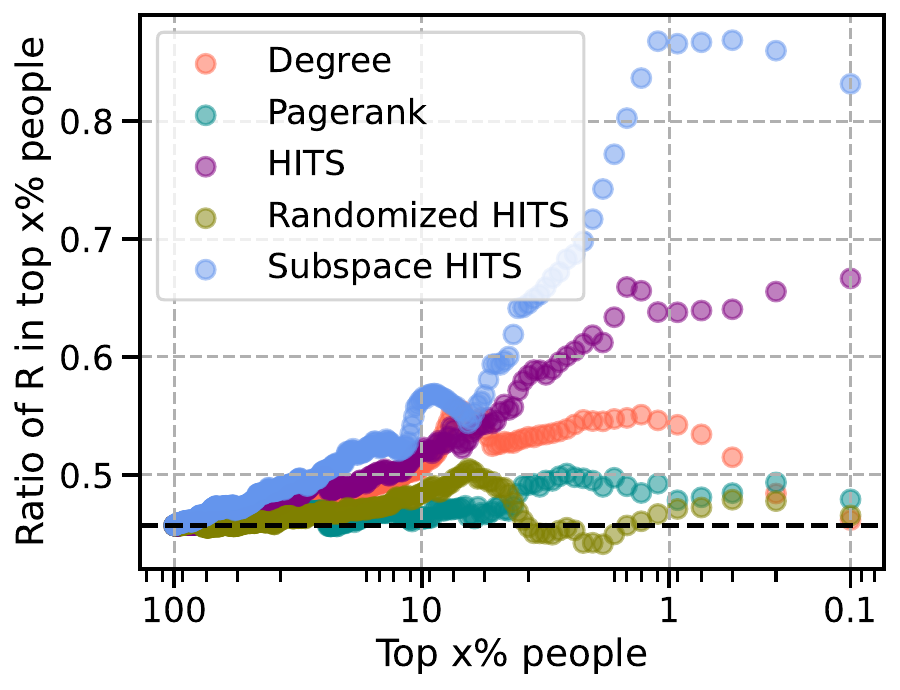}}
\caption{Representation of the minority group R in the ranking of the nodes based on degree (orange), HITS (purple), and PageRank (blue), using $6$ eigenvectors and $f(\lambda) = 1$, for the APS (a), DBLP (b), and Instagram (c) datasets.}
\label{fig:randomizedhits_ranking_alldata}
\end{figure*}

\paragraph{Randomized HITS:} we introduce a random restart probability, similar to the restart probability in PageRank, where with some probability $\epsilon$ a random surfer resets and chooses a node uniformly at random. Conversely, with probability $1-\epsilon$, it adheres to the backward-forward iterations inherent to HITS, yet with a normalization. We choose the restart parameter value to be equal to $\epsilon = 0.15$, just like for the implementation of PageRank in our analysis. This variation has been proposed in Section 5.1 of~\citet{ng2001stable}, formalized as:~\looseness=-1 
    \begin{equation}
    \begin{gathered}
        a^{(t + 1)} = \epsilon \cdot \overrightarrow{\mathbbm{1}} + (1 - \epsilon)\cdot  A_{row}^{T} \cdot h^{(t)}, \\
        h^{(t + 1)} = \epsilon \cdot \overrightarrow{\mathbbm{1}} + (1 - \epsilon)\cdot  A_{col} \cdot a^{(t + 1)}, 
    \end{gathered}
    \label{eq:randomizedhits_original}
    \end{equation}
    where $\overrightarrow{\mathbbm{1}}$ is the all-ones vector, $A_{row}$ and $A_{col}$ are the row- and column-stochastic versions of the adjacency matrix $A$, respectively. We note that randomized HITS is conceptually similar to the SALSA algorithm~\cite{lempel2000stochastic}, for which we find very similar results, omitted in this article due to space constraints.~\looseness=-1

\begin{proposition}
    For a network $G(V,E)$ drawn from BPAM with $N$ nodes, two communities, minority ratio $r$, and homophily parameter $\rho$, the authority scores produced by randomized HITS can be approximated by the indegree distribution, scaled by a coefficient:
    \[ a^{(t)}\approx\epsilon \cdot \overrightarrow{\mathbbm{1}}
+ \textbf{d}^{in} \cdot A(\epsilon,d,t),\]
where $\textbf{d}^{in}$ is the vector of indegrees.~\looseness=-1 
\label{prop:randomizedhitsdegree}
\end{proposition}

We note in the extended analysis that and $A(\epsilon,d,t)$ is bounded above in $t$. This shows that randomized HITS essentially reproduces the degree bias in theory, yet, it is more promising in practice: randomized HITS reproduces degree bias on synthetic data generated from the BPAM with various parameters (Figure~\ref{fig:randomHITS_ranking_symhomophily_directed}), whereas it alleviates bias on real data (Figure~\ref{fig:randomizedhits_ranking_alldata}). Randomized HITS performs similarly to PageRank, except for DBLP, for which PageRank is actually quite similar to the degree ranking, but randomized HITS improves the minority representation as compared to the degree ranking (for example, for DBLP, for the top $10\%$ of the ranks, approximately $16\%$ of them are women in the degree ranking, compared to $21\%$ in randomized HITS, which is close to the population women ratio of $22\%$).~\looseness=-1

\paragraph{Subspace HITS:} Finally, we experimentally explore the potential multiple dimensions in the eigenspace to improve fairness. We combine multiple eigenvectors of the $A^TA$ matrix for computing the authority score, instead of just using the principal eigenvector. The intuition is that more information about different communities may be stored in lower eigenvectors (for example, it has been shown that one extra dimension in PCA greatly improves the fairness of representation of minority groups in the dimensions chosen for projection~\cite{samadi2018price}). We note that this variation has been proposed in Section 5.1 of~\citet{ng2001stable}, formalized as:~\looseness=-1
    \begin{enumerate}
        \item Choose the first $k$ eigenvectors $(v_i)_i$ of $A^TA$ with their corresponding eigenvalues $(\lambda_i)_i$. We choose $k$ between $6$ and $10$ in experiments.
        \item Compute the authority of node $j$ as 
        $a_j = \sum\limits_{i = 1}^{k} f(\lambda_i) (e_j^Tv_i)^2$, where $e_j$ is the $j$-th basis vector and $f(\cdot)$ is a function of our choice (we experiment with $f(\lambda_i) = 1$ and $f(\lambda_i) = \lambda_i^2$).~\looseness=-1
    \end{enumerate}

However, subspace HITS is not consistently fair, as it is sometimes less fair than the degree ranking (for APS and Instagram), and sometimes more fair the degree ranking (for DBLP), in Figure~\ref{fig:randomizedhits_ranking_alldata}. Synthetic data shows a similar behavior, with high homophily showing some improvement over degree, whereas moderate degree mostly reproducing the degree bias (Figure~\ref{fig:randomHITS_ranking_symhomophily_directed}). For the interested reader, we experiment with choosing different number of eigenvectors in the extended version, noting varying levels of fairness. This opens a research path for investigating the optimal number of dimensions and their choice in improving fairness.~\looseness=-1

\section{Conclusions and Future Directions}
\label{sec:conclusions}

In this paper, we provide an in-depth analysis of the structural causes of bias in link analysis ranking. By employing a model of biased networks, we uncover connections between degree bias, homophily, and algorithmic choices. As we formally show the role of homophily in amplifying bias in link analysis ranking algorithms such as HITS, we find nuance and formalism to the previously observed empirical effect called the \emph{tightly knit community} effect.~\looseness=-1

We find several promising directions for future work: first, future work could analyze the behavior of randomized HITS on different network models, given our optimistic empirical results. Second, a future direction could delve deeper into the theory of network embeddings to comprehend the extent of structural bias. Third, it would be interesting to understand the long-term impact of interventions on inequality: how would network dynamics change in response to users viewing more fair rankings, and thus, what is the long-term impact of fairness in link analysis ranking?~\looseness=-1

%% consistent spelling of the heading. We thank the anonymous reviewers for constructive suggestions on improving the manuscript.
% \begin{acks}

\section*{Acknowledgements}
We thank Amin Saberi and Christian Borgs for helpful discussions on this project during the Graph Limits and Processes on Networks program at the Simons Institute. We thank the anonymous reviewers for their constructive suggestions. This work was done in part while A.-A. S. and N.L. were visiting the Simons Institute for the Theory of Computing. N.L. was supported in part by the Netherlands Organisation for Scientific Research (NWO) through Gravitation-grant NETWORKS-024.002.003. This material is based upon work supported by the National Science Foundation under Grant No. 2040971.~\looseness=-1
% \end{acks}

% \bibliographystyle{unsrtnat}
\bibliographystyle{plainnat}

\bibliography{main}

\begin{thebibliography}{47}
\providecommand{\natexlab}[1]{#1}
\providecommand{\url}[1]{\texttt{#1}}
\expandafter\ifx\csname urlstyle\endcsname\relax
  \providecommand{\doi}[1]{doi: #1}\else
  \providecommand{\doi}{doi: \begingroup \urlstyle{rm}\Url}\fi

\bibitem[Agarwal et~al.(2006)Agarwal, Chakrabarti, and
  Aggarwal]{agarwal2006learning}
Alekh Agarwal, Soumen Chakrabarti, and Sunny Aggarwal.
\newblock Learning to rank networked entities.
\newblock In \emph{Proceedings of the 12th ACM SIGKDD international conference
  on Knowledge discovery and data mining}, pages 14--23, 2006.

\bibitem[Angwin et~al.(2016)Angwin, Larson, Mattu, and
  Kirchner]{angwin2016machine}
Julia Angwin, Jeff Larson, Surya Mattu, and Lauren Kirchner.
\newblock Machine bias.
\newblock \emph{ProPublica}, May 2016.

\bibitem[Antunes et~al.(2023)Antunes, Banerjee, Bhamidi, and
  Pipiras]{antunes2023attribute}
Nelson Antunes, Sayan Banerjee, Shankar Bhamidi, and Vladas Pipiras.
\newblock Attribute network models, stochastic approximation, and network
  sampling and ranking algorithms.
\newblock \emph{arXiv preprint arXiv:2304.08565}, 2023.

\bibitem[Avin et~al.(2015)Avin, Keller, Lotker, Mathieu, Peleg, and
  Pignolet]{avin2015homophily}
Chen Avin, Barbara Keller, Zvi Lotker, Claire Mathieu, David Peleg, and
  Yvonne-Anne Pignolet.
\newblock Homophily and the glass ceiling effect in social networks.
\newblock In \emph{Proceedings of the 6th Conference on Innovations in
  Theoretical Computer Science}, pages 41--50, 2015.

\bibitem[Barab{\'a}si and Albert(1999)]{barabasi1999emergence}
Albert-L{\'a}szl{\'o} Barab{\'a}si and R{\'e}ka Albert.
\newblock Emergence of scaling in random networks.
\newblock \emph{science}, 286\penalty0 (5439):\penalty0 509--512, 1999.

\bibitem[Barabasi and Oltvai(2004)]{barabasi2004network}
Albert-Laszlo Barabasi and Zoltan~N Oltvai.
\newblock Network biology: understanding the cell's functional organization.
\newblock \emph{Nature reviews genetics}, 5\penalty0 (2):\penalty0 101--113,
  2004.

\bibitem[Beutel et~al.(2019)Beutel, Chen, Doshi, Qian, Wei, Wu, Heldt, Zhao,
  Hong, Chi, et~al.]{beutel2019fairness}
Alex Beutel, Jilin Chen, Tulsee Doshi, Hai Qian, Li~Wei, Yi~Wu, Lukasz Heldt,
  Zhe Zhao, Lichan Hong, Ed~H Chi, et~al.
\newblock Fairness in recommendation ranking through pairwise comparisons.
\newblock In \emph{Proceedings of the 25th ACM SIGKDD International Conference
  on Knowledge Discovery \& Data Mining}, pages 2212--2220, 2019.

\bibitem[Biega et~al.(2018)Biega, Gummadi, and Weikum]{biega2018equity}
Asia~J Biega, Krishna~P Gummadi, and Gerhard Weikum.
\newblock {Equity of attention: Amortizing individual fairness in rankings}.
\newblock In \emph{Proceedings of the 41st International ACM SIGIR Conference
  on Research \& Development in Information Retrieval}, pages 405--414, 2018.

\bibitem[Borodin et~al.(2005)Borodin, Roberts, Rosenthal, and
  Tsaparas]{Borodin:2005wo}
Allan Borodin, Gareth Roberts, Jeffrey Rosenthal, and Panayiotis Tsaparas.
\newblock {Link analysis ranking: algorithms, theory, and experiments}.
\newblock \emph{Transactions on Internet Technology ({TOIT})}, 5\penalty0 (1),
  February 2005.

\bibitem[Bramoull{\'e} et~al.(2012)Bramoull{\'e}, Currarini, Jackson, Pin, and
  Rogers]{bramoulle2012homophily}
Yann Bramoull{\'e}, Sergio Currarini, Matthew~O Jackson, Paolo Pin, and Brian~W
  Rogers.
\newblock Homophily and long-run integration in social networks.
\newblock \emph{Journal of Economic Theory}, 147\penalty0 (5):\penalty0
  1754--1786, 2012.

\bibitem[Celis et~al.(2017)Celis, Straszak, and Vishnoi]{celis2017ranking}
L~Elisa Celis, Damian Straszak, and Nisheeth~K Vishnoi.
\newblock Ranking with fairness constraints.
\newblock \emph{arXiv preprint arXiv:1704.06840}, 2017.

\bibitem[Cucuringu and Mahoney(2011)]{cucuringu2011localization}
Mihai Cucuringu and Michael~W Mahoney.
\newblock Localization on low-order eigenvectors of data matrices.
\newblock \emph{arXiv preprint arXiv:1109.1355}, 2011.

\bibitem[Cui et~al.(2022)Cui, Mariani, and Medo]{cui2022algorithmic}
Mengtian Cui, Manuel~Sebastian Mariani, and Mat{\'u}{\v{s}} Medo.
\newblock Algorithmic bias amplification via temporal effects: The case of
  pagerank in evolving networks.
\newblock \emph{Communications in Nonlinear Science and Numerical Simulation},
  104:\penalty0 106029, 2022.

\bibitem[Dali and Fortuna(2011)]{dali2011learning}
Lorand Dali and Blaz Fortuna.
\newblock Learning to rank for semantic search.
\newblock \emph{SemSearch@ WWW2011}, 2011.

\bibitem[Dong et~al.(2023)Dong, Ma, Wang, Chen, and Li]{dong2023fairness}
Yushun Dong, Jing Ma, Song Wang, Chen Chen, and Jundong Li.
\newblock Fairness in graph mining: A survey.
\newblock \emph{IEEE Transactions on Knowledge and Data Engineering}, 2023.

\bibitem[Dwork et~al.(2012)Dwork, Hardt, Pitassi, Reingold, and
  Zemel]{dwork2012fairness}
Cynthia Dwork, Moritz Hardt, Toniann Pitassi, Omer Reingold, and Richard Zemel.
\newblock Fairness through awareness.
\newblock In \emph{Proceedings of the 3rd Innovations in Theoretical Computer
  Science Conference}, pages 214--226, 2012.

\bibitem[Esp{\'\i}n-Noboa et~al.(2022)Esp{\'\i}n-Noboa, Wagner, Strohmaier, and
  Karimi]{espin2022inequality}
Lisette Esp{\'\i}n-Noboa, Claudia Wagner, Markus Strohmaier, and Fariba Karimi.
\newblock Inequality and inequity in network-based ranking and recommendation
  algorithms.
\newblock \emph{Scientific reports}, 12\penalty0 (1):\penalty0 1--14, 2022.

\bibitem[Feldman et~al.(2015)Feldman, Friedler, Moeller, Scheidegger, and
  Venkatasubramanian]{Feldman2015}
Michael Feldman, Sorelle~A Friedler, John Moeller, Carlos Scheidegger, and
  Suresh Venkatasubramanian.
\newblock Certifying and removing disparate impact.
\newblock In \emph{proceedings of the 21th ACM SIGKDD international conference
  on Knowledge Discovery and Data Mining}, pages 259--268, 2015.

\bibitem[Fortunato et~al.(2006)Fortunato, Flammini, Menczer, and
  Vespignani]{fortunato2006topical}
Santo Fortunato, Alessandro Flammini, Filippo Menczer, and Alessandro
  Vespignani.
\newblock Topical interests and the mitigation of search engine bias.
\newblock \emph{Proceedings of the National Academy of Sciences}, 103\penalty0
  (34):\penalty0 12684--12689, 2006.

\bibitem[Hardt et~al.(2016)Hardt, Price, and Srebro]{hardt2016equality}
Moritz Hardt, Eric Price, and Nati Srebro.
\newblock Equality of opportunity in supervised learning.
\newblock \emph{Advances in Neural Information Processing Systems}, 29, 2016.

\bibitem[Kleinberg(1999)]{kleinberg1999authoritative}
Jon~M Kleinberg.
\newblock Authoritative sources in a hyperlinked environment.
\newblock \emph{Journal of the ACM (JACM)}, 46\penalty0 (5):\penalty0 604--632,
  1999.

\bibitem[Kleinberg et~al.(2017)Kleinberg, Mullainathan, and
  Raghavan]{Kleinberg:2017gt}
Jon~M Kleinberg, Sendhil Mullainathan, and Manish Raghavan.
\newblock {Inherent Trade-Offs in the Fair Determination of Risk Scores}.
\newblock In \emph{Proceedings of the 8th Innovations in Theoretical Computer
  Science Conference}, volume~67, page~43. Schloss Dagstuhl--Leibniz-Zentrum
  fuer Informatik, 2017.

\bibitem[Lee et~al.(2019)Lee, Karimi, Wagner, Jo, Strohmaier, and
  Galesic]{lee2019homophily}
Eun Lee, Fariba Karimi, Claudia Wagner, Hang-Hyun Jo, Markus Strohmaier, and
  Mirta Galesic.
\newblock Homophily and minority-group size explain perception biases in social
  networks.
\newblock \emph{Nature human behaviour}, 3\penalty0 (10):\penalty0 1078--1087,
  2019.

\bibitem[Lempel and Moran(2000)]{lempel2000stochastic}
Ronny Lempel and Shlomo Moran.
\newblock The stochastic approach for link-structure analysis (salsa) and the
  tkc effect.
\newblock \emph{Computer Networks}, 33\penalty0 (1-6):\penalty0 387--401, 2000.

\bibitem[Ley(2009)]{ley2009dblp}
Michael Ley.
\newblock D{B}{L}{P}: some lessons learned.
\newblock \emph{Proceedings of the VLDB Endowment}, 2\penalty0 (2):\penalty0
  1493--1500, 2009.

\bibitem[Liu et~al.(2009)]{liu2009learning}
Tie-Yan Liu et~al.
\newblock Learning to rank for information retrieval.
\newblock \emph{Foundations and Trends{\textregistered} in Information
  Retrieval}, 3\penalty0 (3):\penalty0 225--331, 2009.

\bibitem[Mariani et~al.(2015)Mariani, Medo, and Zhang]{mariani2015ranking}
Manuel~Sebastian Mariani, Mat{\'u}{\v{s}} Medo, and Yi-Cheng Zhang.
\newblock Ranking nodes in growing networks: When pagerank fails.
\newblock \emph{Scientific reports}, 5\penalty0 (1):\penalty0 1--10, 2015.

\bibitem[McPherson et~al.(2001)McPherson, Smith-Lovin, and
  Cook]{McPherson:2001vg}
Miller McPherson, Lynn Smith-Lovin, and James~M Cook.
\newblock {Birds of a Feather: Homophily in Social Networks}.
\newblock \emph{Annual review of sociology}, 27:\penalty0 415--444, 2001.

\bibitem[Najork et~al.(2007)Najork, Zaragoza, and Taylor]{najork2007hits}
Marc~A Najork, Hugo Zaragoza, and Michael~J Taylor.
\newblock Hits on the web: How does it compare?
\newblock In \emph{Proceedings of the 30th annual international ACM SIGIR
  conference on Research and development in information retrieval}, pages
  471--478, 2007.

\bibitem[Newman and Girvan(2003)]{newman2003mixing}
Mark~EJ Newman and Michelle Girvan.
\newblock Mixing patterns and community structure in networks.
\newblock In \emph{Statistical mechanics of complex networks}, pages 66--87.
  Springer, 2003.

\bibitem[Ng et~al.(2001)Ng, Zheng, and Jordan]{ng2001stable}
Andrew~Y Ng, Alice~X Zheng, and Michael~I Jordan.
\newblock Stable algorithms for link analysis.
\newblock In \emph{Proceedings of the 24th annual international ACM SIGIR
  conference on Research and development in information retrieval}, pages
  258--266, 2001.

\bibitem[Page et~al.(1999)Page, Brin, Motwani, and Winograd]{page1999pagerank}
Lawrence Page, Sergey Brin, Rajeev Motwani, and Terry Winograd.
\newblock The pagerank citation ranking: Bringing order to the web.
\newblock Technical report, Stanford InfoLab, 1999.

\bibitem[Samadi et~al.(2018)Samadi, Tantipongpipat, Morgenstern, Singh, and
  Vempala]{samadi2018price}
Samira Samadi, Uthaipon Tantipongpipat, Jamie~H Morgenstern, Mohit Singh, and
  Santosh Vempala.
\newblock The price of fair {PCA}: One extra dimension.
\newblock \emph{Advances in Neural Information Processing Systems}, 31, 2018.

\bibitem[Singh and Joachims(2019)]{singh2019policy}
Ashudeep Singh and Thorsten Joachims.
\newblock Policy learning for fairness in ranking.
\newblock \emph{Advances in Neural Information Processing Systems}, 32, 2019.

\bibitem[Sinha et~al.(2015)Sinha, Shen, Song, Ma, Eide, Hsu, and
  Wang]{sinha2015overview}
Arnab Sinha, Zhihong Shen, Yang Song, Hao Ma, Darrin Eide, Bo-june~Paul Hsu,
  and Kuansan Wang.
\newblock An overview of {M}icrosoft {A}cademic {S}ervice ({MAS}) and
  applications.
\newblock In \emph{Proceedings of the 24th International Conference on World
  Wide Web}, pages 243--246. ACM, 2015.

\bibitem[Stoica et~al.(2018)Stoica, Riederer, and
  Chaintreau]{stoica2018algorithmic}
Ana-Andreea Stoica, Christopher Riederer, and Augustin Chaintreau.
\newblock Algorithmic glass ceiling in social networks: The effects of social
  recommendations on network diversity.
\newblock In \emph{Proceedings of The Web Conference}, pages 923--932, 2018.

\bibitem[Tang et~al.(2008)Tang, Zhang, Yao, Li, Zhang, and Su]{Tang:08KDD}
Jie Tang, Jing Zhang, Limin Yao, Juanzi Li, Li~Zhang, and Zhong Su.
\newblock Arnetminer: Extraction and mining of academic social networks.
\newblock In \emph{proceedings of the 21th ACM SIGKDD international conference
  on Knowledge Discovery and Data Mining}, pages 990--998, 2008.

\bibitem[Tsioutsiouliklis et~al.(2021)Tsioutsiouliklis, Pitoura, Tsaparas,
  Kleftakis, and Mamoulis]{tsioutsiouliklis2021fairness}
Sotiris Tsioutsiouliklis, Evaggelia Pitoura, Panayiotis Tsaparas, Ilias
  Kleftakis, and Nikos Mamoulis.
\newblock Fairness-aware pagerank.
\newblock In \emph{Proceedings of the Web Conference 2021}, pages 3815--3826,
  2021.

\bibitem[Tsioutsiouliklis et~al.(2022)Tsioutsiouliklis, Pitoura, Semertzidis,
  and Tsaparas]{tsioutsiouliklis2022link}
Sotiris Tsioutsiouliklis, Evaggelia Pitoura, Konstantinos Semertzidis, and
  Panayiotis Tsaparas.
\newblock Link recommendations for pagerank fairness.
\newblock In \emph{Proceedings of the ACM Web Conference 2022}, pages
  3541--3551, 2022.

\bibitem[Veloso et~al.(2008)Veloso, Almeida, Gon{\c{c}}alves, and
  Meira~Jr]{veloso2008learning}
Adriano~A Veloso, Humberto~M Almeida, Marcos~A Gon{\c{c}}alves, and Wagner
  Meira~Jr.
\newblock Learning to rank at query-time using association rules.
\newblock In \emph{Proceedings of the 31st annual international ACM SIGIR
  conference on Research and development in information retrieval}, pages
  267--274, 2008.

\bibitem[Vlasceanu and Amodio(2022)]{vlasceanu2022propagation}
Madalina Vlasceanu and David~M Amodio.
\newblock Propagation of societal gender inequality by internet search
  algorithms.
\newblock \emph{Proceedings of the National Academy of Sciences}, 119\penalty0
  (29):\penalty0 e2204529119, 2022.

\bibitem[Zafar et~al.(2017)Zafar, Valera, Rogriguez, and
  Gummadi]{zafar2017fairness}
Muhammad~Bilal Zafar, Isabel Valera, Manuel~Gomez Rogriguez, and Krishna~P
  Gummadi.
\newblock Fairness constraints: Mechanisms for fair classification.
\newblock In \emph{Artificial Intelligence and Statistics}, pages 962--970.
  PMLR, 2017.

\bibitem[Zehlike and Castillo(2020)]{zehlike2020reducing}
Meike Zehlike and Carlos Castillo.
\newblock Reducing disparate exposure in ranking: A learning to rank approach.
\newblock In \emph{Proceedings of The Web Conference}, pages 2849--2855, 2020.

\bibitem[Zehlike et~al.(2017)Zehlike, Bonchi, Castillo, Hajian, Megahed, and
  Baeza-Yates]{zehlike2017fa}
Meike Zehlike, Francesco Bonchi, Carlos Castillo, Sara Hajian, Mohamed Megahed,
  and Ricardo Baeza-Yates.
\newblock Fa* ir: A fair top-k ranking algorithm.
\newblock In \emph{Proceedings of the 2017 ACM on Conference on Information and
  Knowledge Management}, pages 1569--1578, 2017.

\bibitem[Zehlike et~al.(2022{\natexlab{a}})Zehlike, Yang, and
  Stoyanovich]{zehlike2022fairness}
Meike Zehlike, Ke~Yang, and Julia Stoyanovich.
\newblock Fairness in ranking, part i: Score-based ranking.
\newblock \emph{ACM Computing Surveys}, 55\penalty0 (6):\penalty0 1--36,
  2022{\natexlab{a}}.

\bibitem[Zehlike et~al.(2022{\natexlab{b}})Zehlike, Yang, and
  Stoyanovich]{zehlike2022fairness2}
Meike Zehlike, Ke~Yang, and Julia Stoyanovich.
\newblock Fairness in ranking, part ii: Learning-to-rank and recommender
  systems.
\newblock \emph{ACM Computing Surveys}, 55\penalty0 (6):\penalty0 1--41,
  2022{\natexlab{b}}.

\bibitem[Zhang et~al.(2021)Zhang, Han, Mahajan, Bengani, and
  Chaintreau]{zhang2021chasm}
Yiguang Zhang, Jessy~Xinyi Han, Ilica Mahajan, Priyanjana Bengani, and Augustin
  Chaintreau.
\newblock Chasm in hegemony: Explaining and reproducing disparities in
  homophilous networks.
\newblock \emph{Proceedings of the ACM on Measurement and Analysis of Computing
  Systems}, 5\penalty0 (2):\penalty0 1--38, 2021.

\end{thebibliography}

\newpage

\newpage 
\section{Appendix}

%%%%%%%%%%%%%%%%% BPAM ASYMPTOTICS %%%%%%%%%%%%%%%%%%%%%%%%
\subsection{Asymptotic degree distribution in BPAM} 
\label{subsec:proofssec4}

We start by computing two types of probabilities, for a node $v \in V$ and a color $C \in \{R,B\} $: 

\begin{equation}
    \mathbb{P}(v \mbox{ connects to a node of color } C)
    \label{eq:to}
\end{equation}
and 
\begin{equation}
    \mathbb{P}(v \mbox{ has connections from a node of color } C).
    \label{eq:from}
\end{equation}
Note that the probabilities in equations~\eqref{eq:to} and~\eqref{eq:from} are different. We compute these probabilities asymptotically as the size of the network $N$ grows to infinity. 
Denote by $d_{tot}^{N}(C)$ the total degree of nodes of color $C$. Recall that the total degree in the network of size $N$, is $2Nd$. 
 As in~\citet{avin2015homophily}, denote by $\alpha_N := \frac{d_{tot}(R)}{2Nd}$ (the fraction of edges that connect to or from the red community). It follows from Lemmas $4.4$ and $4.5$ in~\citet{avin2015homophily} that  $\lim\limits_{N\to\infty} \mathbb{E}[\alpha_N]=\alpha < r$, so the fraction of edges connecting to red nodes is smaller than the fraction of red nodes in the network. In~\citet{avin2015homophily} this is called the {\it power inequality}. In fact, as they show, $\alpha_N$ converges in probability to $\alpha$, which is needed for the rest of our argument. 
 Then, compute the probabilities \eqref{eq:to} in the asymptotic regime as:
\begin{align}
    \begin{split}
        p^{out}_{RR} &:= \mathbb{P}(v \mbox{ connects to a node of color } R | v \in R) = \frac{\alpha}{ \alpha + \rho(1-\alpha)}, \\
        p^{out}_{RB} &:= \mathbb{P}(v \mbox{ connects to a node of color } B | v \in R) = \frac{\rho ( 1 - \alpha) }{ \alpha + \rho(1-\alpha)}, \\
        p^{out}_{BR} &:= \mathbb{P}(v \mbox{ connects to a node of color } R | v \in B) = \frac{ \rho \alpha}{ \rho \alpha + 1-\alpha}, \\
        p^{out}_{BB} &:= \mathbb{P}(v \mbox{ connects to a node of color } B | v \in B) = \frac{1 - \alpha}{ \rho \alpha + 1-\alpha},     
    \end{split}
\end{align}
Computing the probabilities in~\eqref{eq:from}  is slightly more complicated: 
    \begin{align}
        \begin{split}
            p^{in}_{BB} &:= \mathbb{P}(v \mbox{ receives connection from  a node of color } B  \,  |\, v \in B) \\
            &= \frac{\frac{(1 - r)(1-\alpha)}{\alpha \rho + 1 - \alpha}}{\frac{r\rho(1-\alpha)}{\alpha + \rho(1 - \alpha)} + \frac{(1 - r)(1-\alpha)}{\alpha \rho + 1 - \alpha}} = \frac{\frac{1 - r}{\alpha \rho + 1 - \alpha}}{\frac{r\rho}{\alpha + \rho(1 - \alpha)} + \frac{1 - r}{\alpha \rho + 1 - \alpha}},\\
            p^{in}_{BR} &:= \mathbb{P}(v \mbox{ receives connection from  a node of color } R\,  |\, v \in B) \\
            &= \frac{\frac{\rho r(1-\alpha)}{\alpha+ \rho (1 - \alpha)}}{\frac{r\rho(1-\alpha)}{\alpha + \rho(1 - \alpha)} + \frac{(1 - r)(1-\alpha)}{\alpha \rho + 1 - \alpha}} = \frac{\frac{\rho r}{\alpha+ \rho (1 - \alpha)}}{\frac{r\rho}{\alpha + \rho(1 - \alpha)} + \frac{1 - r}{\alpha \rho + 1 - \alpha}},\\
            p^{in}_{RR} &:= \mathbb{P}(v \mbox{ receives connection from  a node of color } R  \,  |\, v \in R) \\
            &= \frac{\frac{r\alpha}{\alpha+ \rho (1 - \alpha)}}{\frac{r\alpha}{\alpha + \rho(1 - \alpha)} + \frac{\rho(1 - r)\alpha}{\alpha \rho + 1 - \alpha}} = \frac{\frac{ r}{\alpha+ \rho (1 - \alpha)}}{\frac{r}{\alpha + \rho(1 - \alpha)} + \frac{\rho(1 - r)}{\alpha \rho + 1 - \alpha}},\\
            p^{in}_{RB} &:= \mathbb{P}(v \mbox{ receives connection from  a node of color } B  \,  |\, v \in R) \\
            &= \frac{\frac{\rho(1 - r)\alpha}{\alpha \rho + 1 - \alpha}}{\frac{r\alpha}{\alpha + \rho(1 - \alpha)} + \frac{\rho(1 - r)\alpha}{\alpha \rho + 1 - \alpha}} = \frac{\frac{\rho(1 - r)}{\alpha \rho + 1 - \alpha}}{\frac{r}{\alpha + \rho(1 - \alpha)} + \frac{\rho(1 - r)}{\alpha \rho + 1 - \alpha}}.
        \end{split}
        \label{eq:from-all}
    \end{align}

We continue with a short analysis of the degree distribution of the BPAM. It was proved in~\citet{avin2015homophily} that in the BPAM with two communities, as $N\to\infty$, the limiting degree distribution is a power law distribution with a different exponent for each community. Specifically, denoting by $\mbox{top}_k(C)$ the number of nodes of degree at least $k$ of color $C$, we have:~\looseness=-1
    \begin{align*}
    \begin{split}
        \mbox{top}_k(R) & \sim k^{-\beta_R}, \\
         \mbox{top}_k(R) &\sim k^{-\beta_B},
    \end{split}
    \end{align*}
     where $a \sim b$ means that $a$ is proportional to $b$. Moreover, \citet{avin2015homophily} derive  the closed-form expression for the power law exponents:
\begin{align}
    \begin{split}
        \beta_B &= 1 + \frac{1}{K_B}, \\
        \beta_R &= 1 + \frac{1}{K_R},
    \end{split}
    \label{eq:betas}
\end{align}
where
\begin{equation}
    \begin{gathered}
        K_B = \frac{1}{2} \left( \frac{r\rho}{\alpha + \rho(1 - \alpha)} + \frac{1 - r}{\alpha\rho + 1 - \alpha}\right)>\frac{1}{2}, \\
        K_R = \frac{1}{2} \left( \frac{r}{\alpha + \rho(1 - \alpha)} + \frac{\rho(1 - r)}{\alpha\rho + 1 - \alpha}\right)<\frac{1}{2}.
    \end{gathered}
    \label{eq:cs}
\end{equation}
According to~\citet{avin2015homophily}, it follows from equations~\eqref{eq:betas} and~\eqref{eq:cs} that \[\beta_R > 3 > \beta_B.\] For the proofs of the results in Section~\ref{sec:theory-hits}, we will also need the following two propositions. 

\begin{proposition} 
    For $\beta_B$ defined in equations~\eqref{eq:betas} and~
    \eqref{eq:cs}, it holds that $\beta_B>2$. 
    \label{eq:bpam-betaBineq}
\end{proposition}

\begin{proof}
From equation~\eqref{eq:betas}, we have that $\beta_B > 2$ is equivalent to $K_B < 1$, and from equation~\eqref{eq:cs}, this is equivalent to 
\begin{equation}
    \begin{gathered}
               \frac{r\rho}{\alpha + \rho(1-\alpha)} + \frac{1 - r}{\alpha \rho + 1 - \alpha} <2. 
    \end{gathered}
\label{eq:betabbiggerthan2-1}
\end{equation}
The first fraction in \eqref{eq:betabbiggerthan2-1} is smaller than one because $r<\frac{1}{2}<1-\alpha$. The second fraction is smaller than one because $1-r<1-\alpha$ (recall the power inequality $\alpha<r$). Hence, the total left-hand side of equation~\eqref{eq:betabbiggerthan2-1} is smaller than $2$. This proves the proposition.
\end{proof}

\begin{proposition} We have that  
    \begin{equation}
        \frac{1}{\beta_R-1}>\frac{2}{\beta_B-1}-1,
        \label{eq:bpam-betaB-betaR-ineq0}
    \end{equation}
    or, equivalently,
    \begin{equation}
        2K_B-1<K_R.
        \label{eq:bpam-KB-KR-ineq0}
    \end{equation}
    \label{eq:bpam-betaB-betaR-ineq}
\end{proposition}

\begin{proof} 
We will prove that $4K_B - 2 K_R <2$.    
Substituting the expressions \eqref{eq:cs} for $K_B$ and $K_R$, we have to prove that  
\begin{align*}
    \begin{split}
        &\frac{2r\rho}{\alpha + \rho - \alpha\rho} + \frac{2(1-r)}{\alpha\rho + 1- \alpha} - \frac{\rho(1-r)}{\alpha \rho + 1 - \alpha} - \frac{r}{\alpha + \rho - \alpha\rho}<2\\
        &\Leftrightarrow
        \frac{r}{\alpha + \rho - \alpha\rho} (2\rho - 1)+ \frac{1 - r}{\alpha\rho + 1- \alpha} (2 - \rho) <2.
        \end{split}
        \end{align*}
   Multiplying both sides of the inequality by $(\alpha\rho + 1- \alpha)(\alpha + \rho - \alpha\rho)$, we get 
        \begin{align*}
        \begin{split}
        r(2\rho &-1)(\alpha \rho + 1 - \alpha) + (1-r)(2 - \rho)(\alpha + \rho - \alpha \rho)\\&<2 (\alpha + \rho - \alpha\rho)(\alpha\rho + 1 - \alpha)\\
       &\Leftrightarrow 2r\alpha \rho^2 + 2r\rho - 2r\alpha \rho - r\alpha \rho - r + r\alpha +2\alpha + 2\rho - 2\alpha \rho \\&- 2r\alpha - 2r\rho+2r\alpha\rho - \alpha \rho - \rho^2 + \alpha \rho^2 + r\alpha \rho + r\rho^2 - r\alpha \rho^2\\& < 
        2\alpha^2 \rho + 2\alpha - 2\alpha^2 + 2\alpha \rho^2 + 2\rho - 2\alpha \rho - 2\alpha^2 \rho^2 -2\alpha\rho + 2\alpha^2 \rho\\ &\Leftrightarrow 
        r\alpha\rho^2 - r - r\alpha -\rho^2 +r\rho^2 < 4\alpha^2 \rho - 2\alpha^2 + \alpha \rho^2 -2\alpha^2\rho^2\\ &\Leftrightarrow
        0 < \rho^2 (1+\alpha) + (1-\rho) (r (1 + \alpha)(1 + \rho) - 2\alpha^2 (1 -\rho))
    \end{split}
\end{align*}

Now, we know that $\rho$ and $\alpha$ are non-negative, therefore $ \rho^2 (1+\alpha) \geq 0$. We also know that $\rho \leq 1$, so $ 1 - \rho \geq 0$. We will quickly show that $ r (1 + \alpha)(1 + \rho) - 2\alpha^2 (1 -\rho) \geq 0$. First, we know that $1 > r > \alpha \geq 0$. Therefore, $r(1 + \alpha) \geq 2\alpha^2 \geq 0$. Second, $1 + \rho > 1 - \rho$, concluding the proof.  
%\fi
\end{proof}

%%%%%%%%%%%%%%% HITS APPROXIMATION %%%%%%%%%%%%%%%%%%%%%%%%%%
\subsection{A detailed analysis of HITS} 
\label{sec:approximationproofs}

%%%%%%%%%%%%%%%%%%%%%% NEW VERSION %%%%%%%%%%%%%%%%%%%%%%%%%%

This section details the proof of Theorem~\ref{thm:mainthm}, for which we first focus on proving Propositions~\ref{prop:mf-ineq} and~\ref{prop:homophily_worse_hits_n}, and then tie them together in a mean-field approximation for the authority scores of nodes of different colors. 

As per equation~\eqref{eq:updateequations}, the HITS update equations are defined as follows:
\begin{align}
    \begin{split}
     a^{(t+1)}(v)&=\sum_{w:(w,v)\in E}h^{(t)}(w),\\   
     h^{(t+1)}(v)&=\sum_{w:(v,w)\in E}a^{(t+1)}(w), \forall t \geq 0,
    \end{split}
    \label{eq:hits-def}
\end{align}
starting with $h^{(0)}(v)=1$ for all $v\in V$, and ranking nodes according to the authority scores $a^{(t)}(v)$. 
By iterating equation~\eqref{eq:hits-def} once, we obtain a recursion for $a^{(t+1)}(v)$ in terms of $a^{(t)}(z)$, where $z$ is a vertex connected to $v$ by a backward-forward step $z\leftarrow w \rightarrow v$ (see equation~\eqref{eq:hits-a(t+1)} below). This recursion is central to our analysis, and is split in three terms: 1) $z=v$; 2) $z\ne v$, $z\in B$; 3) $z\ne v$, $z\in R$. Formally, we write:
\begin{align}
    \begin{split}     
     a^{(t+1)}(v)&=\sum_{w:(w,v)\in E}\,\sum_{z:(w,z)\in E}a^{(t)}(z)\\   
     &= \din(v)a^{(t)}(v)\\
     &+ \sum_{C\in\{R,B\}} \sum_{\scriptsize \begin{array}{c} w\in C \\ (w,v)\in E\end{array}}\sum_{\scriptsize \begin{array}{c}z\in B \\ (w,z)\in E\\ z\ne v\end{array}} a^{(t)}(z)\\
    &+ \sum_{C\in\{R,B\}} \sum_{\scriptsize \begin{array}{c} w\in C \\ (w,v)\in E\end{array}}\sum_{\scriptsize \begin{array}{c}z\in R \\ (w,z)\in E,\\ z\ne v\end{array}} a^{(t)}(z), \quad t=1,2,\ldots.
    \end{split}
    \label{eq:hits-a(t+1)}
\end{align}
We will derive a mean-field approximation for equation~\eqref{eq:hits-a(t+1)}. We note that this recursion can also be thought of as path counting for backward-forward paths that start in $v$ (see~\citet{Borodin:2005wo} for an example). 

The first mean-field step is in approximating the fraction of red and blue in- and outneighbors of node $w$, by the corresponding probabilities. 

Let $q_{CC'}$ be the probability that node $v$ of color $C$ has in-edge from node $w$ (of any color), which in turns has out-edge to node $z$ of color $C'$. Then we have
\begin{equation}
    q_{CC'}=p^{in}_{CB}\cdot p^{out}_{BC'}+p^{in}_{CR} \cdot p^{out}_{RC'}.
\end{equation}
Note that
\begin{equation}
\label{eq:q-total}
q_{CB}+q_{CR}=1.
\end{equation}

The second mean-field step is in replacing $a^{(t)}(z)$ in equation~\eqref{eq:hits-a(t+1)} by the average over all vertices of the same color as $z$. This step is justified because our preferential attachment graph can be approximated by its so-called {\it local weak limit}, which is a continuous-time branching process (see the precise  convergence result in~\citet[Theorem 3.5]{antunes2023attribute}). Such processes grow exponentially in time, thus, vertex $w$ in equation~\eqref{eq:hits-a(t+1)} most likely has arrived at the end of the graph formation when most edges were already present. Therefore, in our mean-field approximation we view the directed edge $(w,z)$ as being randomly sampled from the entire graph. Then, the probability that this edge connects to $z$ is proportional to the indegree of $z$. Hence, in the mean-field step, we replace $a^{(t)}(z)$ by its mean with respect to the size-biased distribution of indegrees in color $C$, which we denote by $\overline{a^{(t)}(C)}$: 

\begin{equation}
    \overline{a^{(t)}(C)} = \sum_{z\in C}\frac{\din(z)}{\sum_{u\in C}\din(u)} \cdot a^{(t)}(z), \quad C\in\{R,B\}.
\end{equation}

With this notation, the mean-field approximation of equation~\eqref{eq:hits-a(t+1)} becomes
\begin{align}
\begin{split}
    a^{(t+1)}(v) &\approx \din(v) \cdot a^{(t)}(v)\\& 
     +
    \din(v)(d-1) \cdot q_{CB} \cdot \overline{a^{(t)}(B)}\\& + 
    \din(v)(d-1) \cdot q_{CR} \cdot \overline{a^{(t)}(R)}.
\end{split}    
\label{eq:hits-n-MF}
\end{align}
Here and throughout the paper we  use $\approx$ to denote the mean-field approximation. 

Now, in order to investigate the proportion of the majority (the blue vertices) in the ranking, we will iterate equation~\eqref{eq:hits-n-MF} for $t= 1,2, \ldots$, and approximate its main term when $v$ is blue or red. The analysis relies on the properties of the size-biased moments of the indegree distributions, defined as follows:
\begin{equation}
    \tilde{d}^{in}_t(C)=\frac{1}{\sum_{u\in C}\din(u)}\sum_{u\in C}(\din(u))^t, \quad \forall t \in \mathbb{N}^{*}, C\in \{R,B\}.
\end{equation}

Since in our model outdegree is constant $d$, the indegrees follow a power law distribution with the same exponent as the distribution of total degrees. Therefore, in computations below we will use the following well-known results on power law distribution with exponent $\beta_C>2$:
\begin{align}
\label{eq:EVT}
\begin{split}
\tilde{d}^{in}_t(C)=O_P(1), \quad &\mbox{ if } t< \beta_C-1;\\
\tilde{d}^{in}_t(C)= O_P\left(N^{\frac{t}{\beta_C-1}-1}\right), \quad &\mbox{ if } t> \beta_C-1,
\end{split}
\end{align}
where $O_P(\cdot)$ means that the big-O relation holds in probability.

As before, $|h^{(0)}(v)| =1$ for all $v\in V$. Then, by \eqref{eq:updateequations}, we have
\begin{align}
\begin{split}
\label{eq:hits-1}
    a^{(1)}(v) &=\din(v).
    \end{split}
    \end{align}
    Now, iterating equation \eqref{eq:hits-n-MF} once, we get:
    \begin{align}    
\begin{split}
a^{(2)}(v \in B) &\approx (\din(v))^2 + \din(v)(d-1)\cdot q_{BB} \cdot \tilde{d}^{in}_2(B) \\ &+ 
\din(v)(d-1) \cdot q_{BR}\cdot \tilde{d}^{in}_2(R),
\end{split}
\label{eq:hits-2-B}\\
\begin{split}
a^{(2)}(v \in R) &\approx (\din(v))^2+\din(v)(d-1)\cdot q_{RB} \cdot \tilde{d}^{in}_2(B) \\&+
\din(v)(d-1) \cdot q_{RR} \cdot \tilde{d}^{in}_2(R).
\end{split}
\label{eq:hits-2-R}
\end{align}
Equation~\eqref{eq:hits-1} says that the first iteration of HITS ranks the nodes according to their indegrees. Interestingly, equations \eqref{eq:hits-2-B} and \eqref{eq:hits-2-R} show the enhancement of the majority already in the second iteration. Indeed, since in our model $2>\beta_B-1$, equation~\eqref{eq:EVT} says that the second term in equations~\eqref{eq:hits-2-B}-\eqref{eq:hits-2-R}---the mean-field contribution of the blue nodes---scales as a positive power of $N$, $\tilde{d}^{in}_2(B)=O_P\left(N^{\frac{2}{\beta_B-1}-1}\right)$, while $\tilde{d}^{in}_2(R)=O_P(1)$ because $2<\beta_R-1$. Hence, When a node has a moderate degree, its  approximated HITS score after two iterations is dominated by the $O_P\left(N^{\frac{2}{\beta_B-1}-1}\right)$ term coming from the majority (blue) nodes. Formally, for any vertex $v$ with bounded indegree (of the order $O(1)$), we can write 

\begin{equation}
\begin{gathered}
    a^{(2)}(v \in R) \approx \din(v) \cdot (d-1) \cdot MF^{(2)}(R),\\
    a^{(2)}(v \in B) \approx \din(v) \cdot (d-1) \cdot MF^{(2)}(B),
\end{gathered}
\label{eq:recurrence-hits-factor}
\end{equation}
where 
\begin{equation}
\begin{gathered} 
    MF^{(2)}(R) \approx q_{RB} \cdot \tilde{d}^{in}_2(B),\\
    MF^{(2)}(B) \approx q_{BB} \cdot \tilde{d}^{in}_2(B)
    \end{gathered}
    \label{eq:mf2-approx}
\end{equation}
are what we call the multiplicative factors that allow to compare the HITS ranking with the degree ranking. The following two results show that the blue nodes have gained an advantage, thus proving Propositions~\ref{prop:mf-ineq} and~\ref{prop:homophily_worse_hits_n} from the main text:

\begin{proposition}
    The following hold true: 
    \begin{enumerate}
        \item $MF^{(2)}(B) \geq MF^{(2)}(R)$, \\
        \item Taking $MF^{(2)}(B)$ and $MF^{(2)}(R)$ as functions of the model parameters, the function $F^{(2)}(\rho)=\frac{MF^{(2)}(R,\rho, r)}{MF^{(2)}(B,\rho,r)}$ is increasing in the parameter $\rho$.
    \end{enumerate}
    \label{prop:mf2-proof-appendix}
\end{proposition}

\begin{proof}
The proof of these two results will be the base case for the induction-based proof for Propositions~\ref{prop:mf-ineq} and~\ref{prop:homophily_worse_hits_n}.

To prove the first part, we use equation~\eqref{eq:mf2-approx}:
\begin{equation}
    \begin{gathered}
        MF^{(2)}(B) \geq MF^{(2)}(R) \Leftrightarrow q_{BB}\geq q_{RB} \Leftrightarrow \\ 
        p_{BB}^{in} \cdot p_{BB}^{out} + p_{BR}^{in} \cdot p_{RB}^{out} \geq p_{RB}^{in} \cdot p_{BB}^{out} + p_{RR}^{in} \cdot p_{RB}^{out} \Leftrightarrow \\ 
        \left( p_{BB}^{in} - p_{RB}^{in}\right) \cdot p_{BB}^{out} + \left(p_{BR}^{in} - p_{RR}^{in} \right) \cdot p_{RB}^{out} \geq 0.
    \end{gathered}
    \label{eq:mf2bmf2r-proof1}
\end{equation}

We know from equation~\eqref{eq:from-all} that $p_{BB}^{in} + p_{BR}^{in} = 1$ and $p_{RB}^{in} + p_{RR}^{in} = 1$, and so equation~\eqref{eq:mf2bmf2r-proof1} is equivalent to 

\begin{equation}
    \begin{gathered}
        \left( p_{BB}^{in} - p_{RB}^{in}\right) \cdot \left(p_{BB}^{out} - p_{RB}^{out}\right) \geq 0
    \end{gathered}
    \label{eq:mf2bmf2r-proof2}
\end{equation}

We compute 

\begin{equation}
    \begin{gathered}
        p_{BB}^{out} - p_{RB}^{out} = \frac{(1-\rho^2)\alpha(1-\alpha)}{(\alpha\rho + 1 - \alpha)(\alpha + \rho(1-\alpha))}
    \end{gathered}
\end{equation}
and
\begin{equation}
    \begin{gathered}
        p_{BB}^{in} - p_{RB}^{in} = \frac{r(1-r)(\alpha + \rho(1-\alpha))(1-\rho^2)}{(r\rho u_{\alpha,\rho} + (1-r)u_{1-\alpha,\rho})(r u_{\alpha,\rho} + \rho(1-r)u_{1-\alpha,\rho})}
    \end{gathered}
\end{equation}
where, for space brevity, we denoted by $u_{\alpha,\rho} := \alpha\rho + 1 - \alpha$ and by $u_{1-\alpha,\rho} := \rho(1-\alpha) + \alpha$.

Therefore, equation~\eqref{eq:mf2bmf2r-proof2} is equivalent to 

\begin{equation}
    \begin{gathered}
        \frac{r(1-r)\alpha(1-\alpha)(1-\rho^2)^2}{(r\rho u_{\alpha,\rho}+ (1-r)u_{1-\alpha,\rho})(r u_{\alpha,\rho} + \rho(1-r)u_{1-\alpha,\rho})}.
    \end{gathered}
\end{equation}

Since $\alpha ,r$, and $\rho$ are smaller than $1$, both the numerator and the denominator are positive. 

To prove the second part, we take the expressions $q_{CC'}, p_{CC'}^{in}$, and $p_{CC'}^{out}$ for colors $C,C' \in \{R,B \}$ as functions of $\rho$, it is enough to show that $\frac{q_{RB}(\rho)}{q_{BB}(\rho)}$ is an increasing function in $\rho$. Writing out the closed-form formulas and simplifying, this reduces to showing that 

\begin{equation}
    \begin{gathered}
        \frac{\alpha}{1 - \alpha} \cdot \frac{1 - (2\alpha - r)}{2\alpha - r} \cdot \frac{\rho (1-r) (\alpha + \rho - \alpha \rho)^2 + \rho r (\alpha \rho + 1 - \alpha)^2}{ (1-r) (\alpha + \rho - \alpha \rho)^2 + \rho^2 r (\alpha \rho + 1 - \alpha)^2}
    \end{gathered}
\end{equation}
is an increasing function in $\rho$. This is easily seen by differentiating with respect to $\rho$. Note: in simplifying, we have also used the fact that $\alpha$ is the fixed point of a function 
\begin{equation}
    A(\alpha) = \frac{1}{2} \left(r + \frac{r\alpha}{\alpha + \rho - \alpha \rho} + \frac{\alpha \rho (1-r)}{\alpha \rho + 1 - \alpha}\right)
\end{equation}
and that $1 - \alpha$ is the fixed point of a function 
\begin{equation}
B(\alpha) = \frac{1}{2} \left(1 - r + \frac{(1-r)(1-\alpha)}{\alpha \rho + 1 - \alpha} + \frac{r \rho (1-\alpha)}{\alpha + \rho  - \alpha\rho}\right).
\end{equation}
These fixed points follow from the analysis in~\citet{avin2015homophily}, knowing that asymptotically, the fraction of edges towards the red population (which is $\alpha$) and the fraction of edges towards the blue population (which is $1 - \alpha$), are at equilibrium. A numerical illustration of $F^{(2)}(\rho)$ as a function of $\rho$ can be found in Figure~\ref{fig:f2rho_numerical}. (As we will see further on, the function $F^{(t)}(\rho)$ will be the same for $t > 2$ as well.) Finally, it is easy to notice that when $\rho = 1$, most terms simplify and we obtain $F^{(2)}(1) = 1$.

\begin{figure}
\centering
\includegraphics[width=0.35\textwidth] {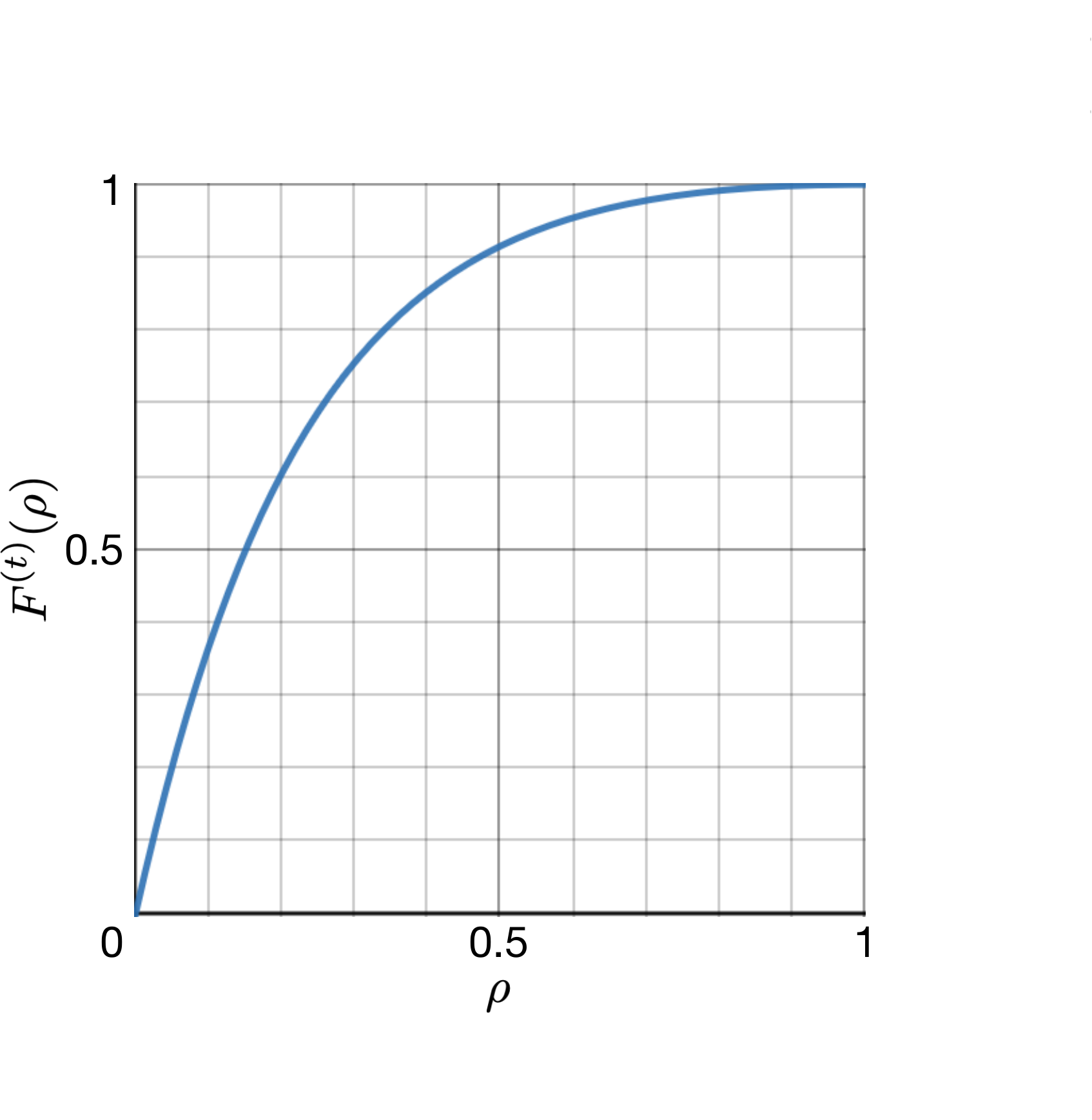}
\caption{Numerical illustration of $F^{(t)}(\rho)$ as a function of $\rho$, for $t = 2$. }
\label{fig:f2rho_numerical}
\end{figure}
\end{proof}

Moving on to a short analysis of nodes of top degree, according to the properties of power laws, the top degrees of red nodes are of order $O_P\left(N^{\frac{1}{\beta_R-1}}\right)$. We know from Proposition~\eqref{eq:bpam-betaB-betaR-ineq} that
\begin{equation}
\label{eq:hits-2-degree-term-order}
\frac{1}{\beta_R-1}>\frac{2}{\beta_B-1}-1,
\end{equation}
so after the second iteration, HITS should rank top-degree red nodes generally higher than mediocre blue nodes. 

As we continue with the third iteration of HITS by \eqref{eq:hits-n-MF}, we must compute $\overline{a^{(2)}(B)}$ and $\overline{a^{(2)}(R)}$, which are the size-biased averages of equations~\eqref{eq:hits-2-B} and~\eqref{eq:hits-2-R}. We obtain: 
\begin{align}
\begin{split}
\overline{a^{(2)}(B)} &= \tilde{d}^{in}_3(B) +\left(\tilde{d}^{in}_2(B)\right)^2(d-1)\cdot q_{BB} \\
& + \tilde{d}^{in}_2(B) \cdot \tilde{d}^{in}_2(R) \cdot (d-1)\cdot q_{BR} \\
&= O_P\left(N^{\frac{3}{\beta_B-1}-1}\right) + O_P\left(N^{\frac{4}{\beta_B-1}-2}\right)\\& +  O_P\left(N^{\frac{2}{\beta_B - 1} - 1}\right) \cdot O_P(1) \\
&= O_P\left(N^{\frac{3}{\beta_B-1}-1}\right),
\label{eq:mf-2-B}
\end{split}
\end{align}
\begin{align}
\begin{split}
\overline{a^{(2)}(R)} &= \tilde{d}^{in}_3(R) + \tilde{d}^{in}_2(R)\cdot \tilde{d}^{in}_2(B) \cdot (d-1)\cdot q_{RB} \\
&+ \left(\tilde{d}^{in}_2(R)\right)^2d(d-1)\cdot q_{RR} \\  
&= O_P\left(N^{\max\{0,\frac{3}{\beta_R-1}-1\}}\right) + O_P(1) \cdot O_P\left( N^{\frac{2}{\beta_B - 1} - 1} \right)\\& + O_P\left(1\right) \\
&= o_P\left(N^{\frac{3}{\beta_B-1}-1}\right).
\end{split}
\label{eq:mf-2-R}
\end{align}
We thus note that the term $\tilde{d}^{in}_{3}(B)$ dominates $\overline{a^{(2)}(B)}$, and it is of a higher order of magnitude than $\overline{a^{(2)}(R)}$. Looking back at equation~\eqref{eq:hits-n-MF}, we see that this term has a contribution of $q_{BB}$ in $a^{(3)}(v\in B)$ and a contribution of $q_{RB}$ in $a^{(3)}(v\in R)$. By the proof of Proposition~\ref{prop:mf2-proof-appendix} we know that $q_{BB} \geq q_{RB}$. To see exactly the closed-form approximation of this term in the HITS score of the two communities, we replace equations~\eqref{eq:mf-2-B} and~\eqref{eq:mf-2-R} in equation~\eqref{eq:hits-n-MF} for $t = 3$, obtaining

\begin{align*}
    \begin{split}
        a^{(3)}&(v \in B) = \left( \din(v)\right)^3  + \left( \din(v)\right)^2  (d-1) \cdot q_{BB} \cdot \tilde{d}^{in}_{2}(B) \\
        &+\left( \din(v)\right)^2  (d-1) \cdot q_{RB} \cdot \tilde{d}^{in}_{2}(R)  \\
        &+\din(v)  (d-1) \cdot q_{BB} \cdot \tilde{d}^{in}_{3}(B) + \din(v)  (d-1)^3 \cdot q_{BB}^2 \cdot \left(\tilde{d}^{in}_{2}(B) \right)^2 \\
        &+\din(v)  (d-1)^3 \cdot q_{BB} \cdot q_{BR} \cdot \tilde{d}^{in}_{2}(R) \cdot \tilde{d}^{in}_{2}(B) \\ 
        &+\din(v)  (d-1) \cdot q_{BR} \cdot \tilde{d}^{in}_{3}(R) \\
        &+ \din(v)  (d-1)^3 \cdot q_{BR} \cdot q_{RR} \cdot \left(\tilde{d}^{in}_{2}(R) \right)^2 \\
        &+\din(v)  (d-1)^3 \cdot q_{BR} \cdot q_{RB} \cdot \tilde{d}^{in}_{2}(R) \cdot \tilde{d}^{in}_{2}(B), \\ 
        a^{(3)}&(v \in R)= \left( \din(v)\right)^3 d + \left( \din(v)\right)^2  (d-1) \cdot q_{RB} \cdot \tilde{d}^{in}_{2}(B) \\
        &+\left( \din(v)\right)^2  (d-1) \cdot q_{RR} \cdot \tilde{d}^{in}_{2}(R) \\
        &+\din(v)  (d-1) \cdot q_{RB} \cdot \tilde{d}^{in}_{3}(B) \\
        &+ \din(v)  (d-1)^3 \cdot q_{RB} \cdot q_{BB} \cdot \left(\tilde{d}^{in}_{2}(B) \right)^2 \\
        &+\din(v)  (d-1)^3 \cdot q_{RB} \cdot q_{BR} \cdot \tilde{d}^{in}_{2}(R) \cdot \tilde{d}^{in}_{2}(B)  \\ 
        &+\din(v)  (d-1) \cdot q_{RR} \cdot \tilde{d}^{in}_{3}(R) + \din(v)  (d-1)^3 \cdot  q_{RR}^2 \cdot \left(\tilde{d}^{in}_{2}(R) \right)^2 \\
        &+\din(v)  (d-1)^3 \cdot q_{RR} \cdot q_{RB} \cdot \tilde{d}^{in}_{2}(R) \cdot \tilde{d}^{in}_{2}(B).
    \end{split}
\end{align*}

Clearly, when $\din(v)=O(1)$, the term $\tilde{d}^{in}_{3}(B)$ dominates both $a^{(3)}(v \in R)$ and $a^{(3)}(v \in B)|$, so we can write 

\begin{equation}
\begin{gathered}
    a^{(3)}(v \in R) \approx \din(v) \cdot (d-1) \cdot MF^{(3)}(R),\\
    a^{(3)}(v \in B) \approx \din(v) \cdot (d-1) \cdot MF^{(3)}(B),
\end{gathered}
\label{eq:recurrence-hits-factor3}
\end{equation}
where 

\begin{equation}
\begin{gathered} 
    MF^{(3)}(R) \approx q_{RB} \cdot \tilde{d}^{in}_3(B),\\
    MF^{(3)}(B) \approx q_{BB} \cdot \tilde{d}^{in}_3(B)
    \end{gathered}
\end{equation}
are the multiplicative factors for $t = 3$. Thus, by the exact same proof, Proposition~\ref{prop:mf2-proof-appendix} can be proved for $t = 3$. A simple inductive argument will show that in all subsequent iterations of \eqref{eq:hits-a(t+1)}, the term $\tilde{d}^{in}_{t}(B)$ of the order $O_P\left(N^{\frac{t}{\beta_B-1}-1}\right)$, dominates $\overline{a^{(t)}(R)}$ and $\overline{a^{(t)}(B)}$, and when $\din(v)=O(1)$, we can compute its coefficient in the same way as we did before: 
\begin{equation}
\begin{gathered}
    a^{(t)}(v \in R) \approx \din(v) \cdot (d-1) \cdot MF^{(t)}(R),\\
    a^{(t)}(v \in B) \approx \din(v) \cdot (d-1) \cdot MF^{(t)}(B),
\end{gathered}
\label{eq:recurrence-hits-factor-form}
\end{equation}
where 

\begin{equation}
\begin{gathered} 
    MF^{(t)}(R) \approx q_{RB} \cdot \tilde{d}^{in}_t(B),\\
    MF^{(t)}(B) \approx q_{BB} \cdot \tilde{d}^{in}_t(B)
    \end{gathered}
\end{equation}
As we have seen before, it is now no different to generalize Proposition~\ref{prop:mf2-proof-appendix} for any $t > 2$. 

% this is the last step of the proof, the induction: 
The induction involves a few properties stemming from the recursion equations and the approximation used. We start by looking at the majority community $B$ (arguing that a similar argument goes through for community $R$):

\begin{enumerate}
    \item $a^{(t)}(v \in B)$ is a polynomial in $d^{in}(v)$ of degree $t$, with a leading coefficient equal to $d$; 
    \item The term $d_{t}^{in}(B)$ is the dominant term (asymptotically in $N$) in $a^{(t)}(v \in B)$, as a coefficient to $d^{in}(v)$.
\end{enumerate}

The first point is easy to see by induction, knowing our base case from equations~\eqref{eq:hits-2-B} and~\eqref{eq:hits-2-R} and the recursion equation~\eqref{eq:hits-n-MF} (which is a linear equation in $d^{in}(v)$). For the second point, we'll use the mean-field equations and the induction hypothesis (assumed true for $t-1$ with the goal of showing it for $t$). When computing $\overline{a^{(t - 1)}(B)}$, we essentially average $a^{(t - 1)}(z)$ over all $z \in B$ with respect to the size-biased distribution of $\din(z\in B)$. Since from the induction hypothesis we know that $a^{(t - 1)}(v \in B)$ is a polynomial in $d^{in}(v)$ of degree $t - 1$ with a leading coefficient equal to $d$, we will show that the dominant term of $\overline{a^{(t - 1)}(B)}$ is $d_t^{in}(B)$. First of all, clearly this term exists with coefficient $d$ from our previous remark. Secondly, we need to show that it is in fact the dominant term. Since from the induction hypothesis the term $d_{t - 1}^{in}(B)$ is the dominant term (asymptotically in $N$) in $a^{(t - 1)}(v \in B)$, as a coefficient to $d^{in}(v)$, this term will turn into $d_{t - 1}^{in}(B) \cdot d_{2}^{in}(B)$ in the averaging process (with some coefficient). Now, in comparing $d_{t - 1}^{in}(B) \cdot d_{2}^{in}(B)$ and $d_t^{in}(B)$, $d_t^{in}(B)$ clearly dominates, since

\begin{equation}
    \frac{t}{\beta_B - 1}  - 1 > \frac{t - 1 }{\beta_B - 1} - 1 + \frac{2 }{\beta_B - 1} - 1 \Leftrightarrow \beta_B > 2,
\end{equation}
which we know to be true. Finally, no other term in $\overline{a^{(t - 1)}(B)}$ is competitive by the same argument. By a similar argument, $d_t^{in}(R)$ is the dominant term in $\overline{a^{(t - 1)}(R)}$. Thus, $d_t^{in}(B)$ is the dominant term in $a^{(t)}(v \in B)$, as the summand $a^{(t - 1)}(B)$ in the approximation contributes $d_{t -1}^{in}(B)$ as a dominant term (clearly dominated), the summand $\overline{a^{(t - 1)}(B)}$ contributes $d_{t}^{in}(B)$ as a dominant term, and the summand $\overline{a^{(t - 1)}(R)}$ contributes, as a dominant term, the larger order term between $d_{t}^{in}(R)$ (clearly dominated as $\beta_B < \beta_R$) and $d_{t-1}^{in}(B)$. Thus, this shows the second part, 
that the term $d_{t}^{in}(B)$ is the dominant term (asymptotically in $N$) in $a^{(t)}(v \in B)$, as a coefficient to $d^{in}(v)$ (coming from $d^{in}(v)$ multiplied by $\overline{a^{(t - 1)}(B)}$ in the mean-field approximation). 

We finalize our analysis by looking at the top red vertices with degree $O_P\left(N^{\frac{1}{\beta_R-1}}\right)$. In iteration~$t>2$, the largest contribution of their degree is $O_P\left(N^{\frac{t}{\beta_R-1}}\right)$, while the largest competing term comes from the mean-field contribution of the blue vertices in $a^{(t - 1)}(v)$, so this term is $(\din(v))^2 \cdot \overline{a^{(t-2)}(B)}=O_P\left(N^{\frac{2}{\beta_R-1}+\frac{t-1}{\beta_B-1}-1}\right)$. When $t=3$, the degree term is still of a larger order of magnitude due to \eqref{eq:hits-2-degree-term-order}. However, in subsequent iterations, since \[\frac{1}{\beta_B-1}>\frac{1}{\beta_R-1},\]
the mean-field contribution of the blue vertices grows faster, has an increasing share of the backward-forward-paths, and at the same time, contributes with a smaller factor $q_{BR} \leq q_{BB}$. This explains the fact that the majority vertices are increasingly enhanced in subsequent iterations of HITS, as the backward-forward paths counted in HITS become longer. This concludes the proof of Theorem~\ref{thm:mainthm}.~\looseness=-1

% \pagebreak

\subsection{A detailed analysis of randomized HITS}
\label{sec:randomizedhitsdegreeargument}

In this section, we detail a short argument for why randomized HITS closely follows the degree ranking in BPAM. We recall the iterative process defining randomized HITS from equation~\eqref{eq:randomizedhits_original}, which we transpose and rewrite for ease of notation (noting that $a(\cdot)$ and $h(\cdot)$ are row vectors and now $ \overrightarrow{\mathbbm{1}}$ defines the row of ones):~\looseness=-1 
    \begin{equation}
    \begin{gathered}
        a^{(t + 1)} = \epsilon \cdot \overrightarrow{\mathbbm{1}} + (1 - \epsilon)\cdot h^{(t)} \cdot A_{row} , \\
        h^{(t + 1)} = \epsilon \cdot \overrightarrow{\mathbbm{1}} + (1 - \epsilon) \cdot a^{(t + 1)} \cdot  A_{col}^T , 
    \end{gathered}
    \label{eq:recursion_randomizedhits}
    \end{equation}
    where $\overrightarrow{\mathbbm{1}}$ is the all-ones vector, $A_{row}$ and $A_{col}$ are the row- and column-stochastic versions of the adjacency matrix $A$, respectively.

We quickly note that $A_{row}$ is essentially normalizing the adjacency matrix by the constant outdegree of the BPAM. Equivalently, $A_{row}$ is the transition matrix for taking a `forward' step from a node, following their outdegree. Similarly, $A_{col}^T$ is essentially normalizing the adjacency matrix by the indegree of the BPAM, equivalent to being the transition matrix for taking a `backward' step from a node, following their indegree. Iterating the recursion of authority scores from equation~\eqref{eq:recursion_randomizedhits}, we get: 

\begin{equation}
    \begin{gathered}
    a^{(t + 1)} = \epsilon \cdot \overrightarrow{\mathbbm{1}} + \epsilon (1 - \epsilon) \cdot \overrightarrow{\mathbbm{1}} \cdot A_{row}  +  (1 - \epsilon)^2  \cdot a^{(t)} \cdot A_{col}^T A_{row} 
    \end{gathered}
    \label{eq:randomizedhits_authrecursion}
\end{equation}

Equation~\eqref{eq:randomizedhits_authrecursion} is similar to the Pagerank equation~\eqref{eq:pagerank}, with transition matrix $A_{col}^TA_{row}$ (the backward-forward matrix) and damping factor $(1-\epsilon)^2$. We observe that 
\begin{align}
\label{eq:randomized-hits-dout}
    \left(\overrightarrow{\mathbbm{1}} \cdot A_{row}\right)(v) =  \sum\limits_{w =1}^{N}\left( A_{row}\right)_{wv} = \sum\limits_{w: (w,v)\in E} \frac{1}{\dout(w)}.
\end{align}
Notice that in our BPAM with constant outdegree $d$, we get
\begin{equation}
\begin{gathered}
    \overrightarrow{\mathbbm{1}} \cdot A_{row} = \frac{1}{d}\, \textbf{d}^{in},
\end{gathered}
\end{equation}
where $\textbf{d}^{in}$ is the row vector of indegrees \begin{equation}
        \textbf{d}^{in} =  \left(d_1^{in}, d_2^{in}, \cdots, d_N^{in} \right). 
\end{equation}
Iterating equation~\eqref{eq:randomizedhits_authrecursion} over $t$ in BPAM, we obtain:
\begin{align}
    \begin{split}
        a^{(t + 1)} &= (1 - \epsilon)^{2t} \cdot a^{(1)} \cdot \left(A_{col}^TA_{row} \right)^t \\
        &+ \left( \frac{\epsilon(1-\epsilon) }{d} \cdot \textbf{d}^{in} + \epsilon \cdot \overrightarrow{\mathbbm{1}}\right) \cdot \sum\limits_{k = 0}^{t - 1} (1 - \epsilon)^{2k} \left(A_{col}^TA_{row} \right)^k.
    \end{split}
    \label{eq:iteration-stochastic-hits}
\end{align}
We note that 
\begin{equation}
    a^{(1)} = \frac{1 - \epsilon}{d} \cdot \textbf{d}^{in} + \epsilon \cdot \overrightarrow{\mathbbm{1}}.
\end{equation}
Substituting this into \eqref{eq:iteration-stochastic-hits} and writing separately the term for $k=0$, we get
\begin{equation}
    \begin{gathered}
        a^{(t + 1)} = (1 - \epsilon)^{2t} \cdot \left( \frac{1 - \epsilon}{d} \cdot \textbf{d}^{in} + \epsilon \cdot \overrightarrow{\mathbbm{1}} \right) \cdot \left(A_{col}^TA_{row} \right)^t \\
        +\frac{\epsilon(1-\epsilon)}{d} \cdot  \textbf{d}^{in} + \epsilon \cdot \overrightarrow{\mathbbm{1}}\\
        + \left( \frac{\epsilon(1-\epsilon)}{d} \cdot  \textbf{d}^{in} + \epsilon \cdot \overrightarrow{\mathbbm{1}}\right) \cdot \sum\limits_{k = 1}^{t - 1} (1 - \epsilon)^{2k} \left(A_{col}^TA_{row} \right)^k.
    \end{gathered}
    \label{eq:randomized-hits-at+1}
\end{equation}
To get more insight into further iterations, we
compute 
\[\left(A^T_{col}A_{row}\right)_{zv}=\sum_{w=1}^N
\frac{A_{wz}}{\din(z)} \cdot \frac{A_{wv}}{d}.\]
Now, we approximate $A_{wv}$ by its average conditioned on degrees: 
\[A_{wv}\approx \frac{\din(v)d}{Nd}=\frac{\din(v)}{N}.\]
Substituting this, we derive
\begin{align*}
\begin{split}
    &\left(\overrightarrow{\mathbbm{1}}\cdot A_{col}^TA_{row}\right)(v)=\sum_{z=1}^N1\cdot\left(\sum_{w=1}^N\frac{A_{wz}}{\din(z)} \cdot \frac{A_{wv}}{d}\right)\\    &\approx\sum_{w=1}^N\sum_{z=1}^N\frac{\din(z)}{\din(z)N}\cdot \frac{\din(v)}{Nd}=\frac{\din(v)}{d}. 
\end{split}
\end{align*}
Also, we notice that in further iterations,  
\begin{align*}
\begin{split}
    &\left(\textbf{d}^{in}\cdot A_{col}^TA_{row} \right)(v)=\sum_{z=1}^N\din(z)\cdot\left(\sum_{w=1}^N\frac{A_{wz}}{\din(z)} \cdot \frac{A_{wv}}{d}\right)\\    &\approx\sum_{w=1}^N\sum_{z=1}^N\din(z)\,\frac{\din(z)}{\din(z)N} \cdot \frac{\din(v)}{Nd}=\din(v). 
\end{split}
\end{align*}
From this and equation \eqref{eq:randomized-hits-at+1}, we see that in the mean-field approximation, 
$a^{(t+1)}$ has the term  $\epsilon \cdot \overrightarrow{\mathbbm{1}}$, and the rest of the terms proportional to $\textbf{d}^{in}$. We conclude that in BPAM, randomized HITS ranks approximately by the indegree.

In a real-world dataset, the outdegrees are not constant, so, for instance, $a^{(1)}(\cdot)$ ranks nodes by the right-hand side of equation~\eqref{eq:randomized-hits-dout}, and all subsequent iterations will have such term as well. Then, like in PageRank, it is beneficial to receive edges from nodes of small outdegree, but unlike in PageRank, the contribution of high indegree neighbors is counterbalanced thanks to the division by indegrees in $A^T_{col}$. If minority nodes tend to receive edges from nodes of lower outdegree, then they will benefit in the ranking produced, achieving higher ranks.

%%%%%%%%%%%%%% subspace hits: more eigenvects %%%%%%%%%%%%%%%%%%%%%%%%%
\subsection{Subspace HITS: an analysis of various eigenvectors}
\label{sec:multipleeigenvectorsexp}
\begin{figure*}[!ht]
\centering
\subfloat[APS, $f(\lambda) = 1$]
{\includegraphics[width=0.3\textwidth] {figures/APS_randomized_subspace_pev_prmanual_degree_glassceiling_avg_final_percplottingxaxis_sns_xlog_nolegend.pdf}}
\subfloat[APS, $f(\lambda) = 1$, multiple EVs]
{\includegraphics[width=0.3\textwidth] {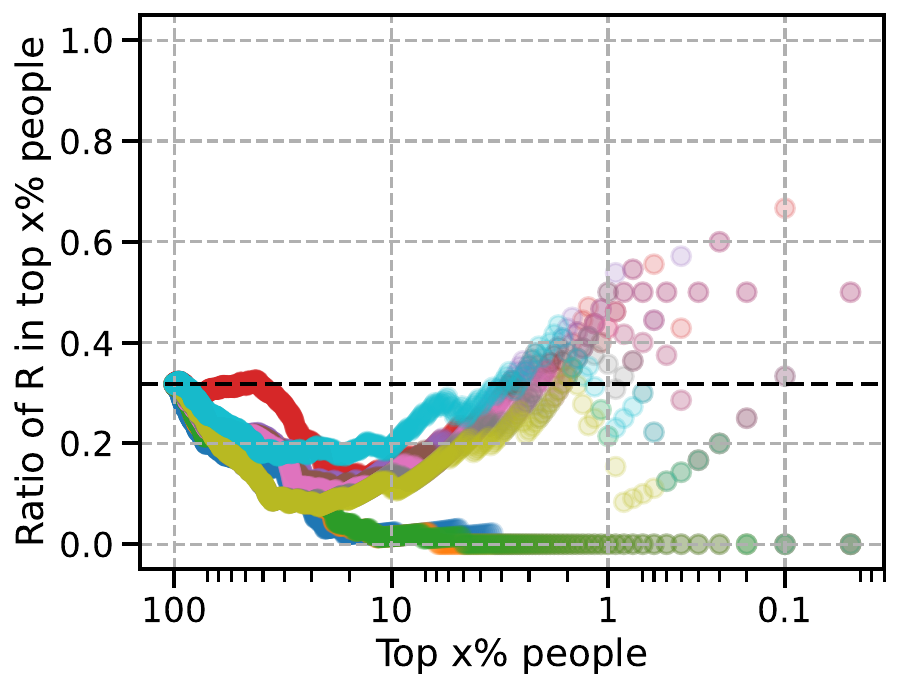}}
\subfloat[APS, $f(\lambda) = \lambda^2$, multiple EVs]
{\includegraphics[width=0.3\textwidth] {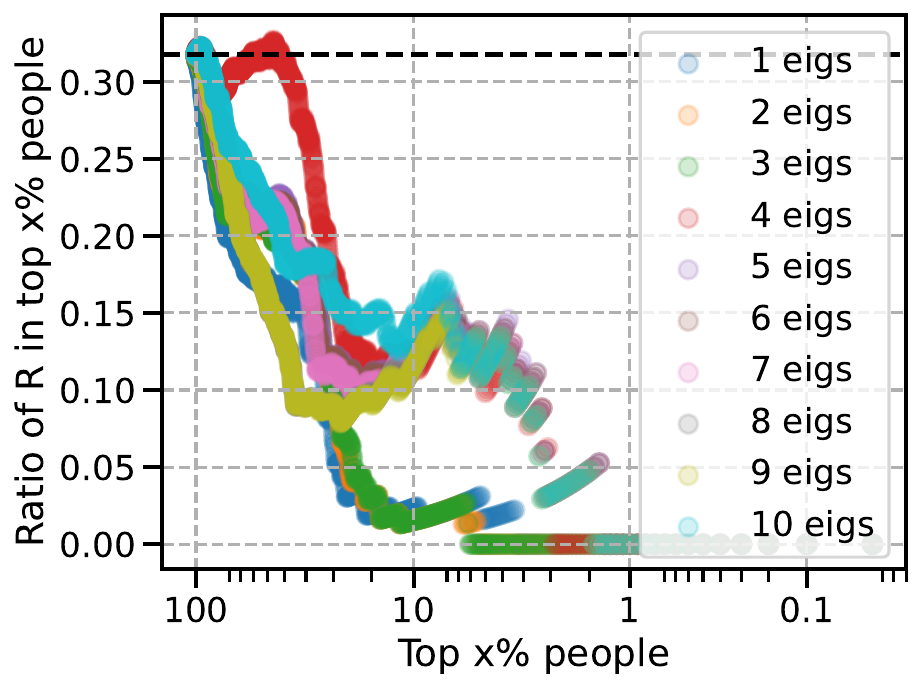}}
\caption{Representation of the minority group R in the ranking of the nodes based on degree (orange), HITS (purple), and Pagerank (blue), for the APS dataset, using $6$ eigenvectors and $f(\lambda) = 1$ (a). Figures (b) and (c) show the representation of the minority group when using a varied number of eigenvectors for $f(\lambda) = 1$ (b) and $f(\lambda) = \lambda^2$ (c).}
\label{fig:randomizedhits_ranking_APS}
\end{figure*}

\begin{figure*}[!ht]
\centering
\subfloat[DBLP, $f(\lambda) = 1$]
{\includegraphics[width=0.3\textwidth] {figures/DBLPv10_randomized_subspace_pev_prmanual_degree_glassceiling_avg_final_percplottingxaxis_sns_xlog_nolegend.pdf}}
\subfloat[DBLP, $f(\lambda) = 1$, multiple EVs]
{\includegraphics[width=0.3\textwidth] {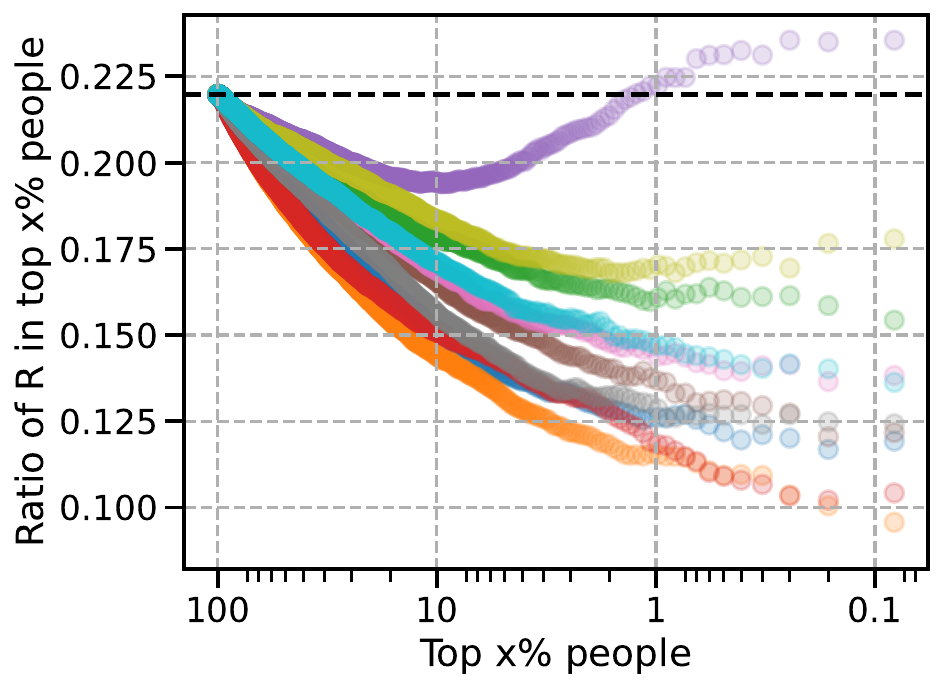}}
\subfloat[DBLP, $f(\lambda) = \lambda^2$, multiple EVs]
{\includegraphics[width=0.3\textwidth] {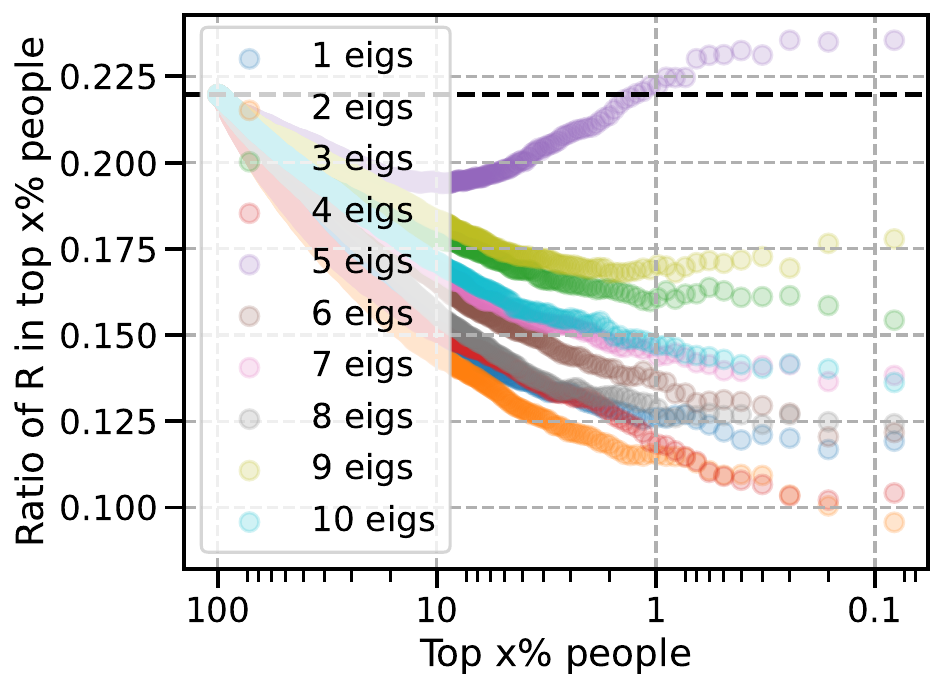}}
\caption{Representation of the minority group R in the ranking of the nodes based on degree (orange), HITS (purple), and Pagerank (blue), for the DBLP dataset, using $10$ eigenvectors and $f(\lambda) = 1$ (a). Figures (b) and (c) show the representation of the minority group when using a varied number of eigenvectors for $f(\lambda) = 1$ (b) and $f(\lambda) = \lambda^2$ (c).}
\label{fig:randomizedhits_ranking_dblpv10}
\end{figure*}

\begin{figure*}[!ht]
\centering
\subfloat[Instagram, $f(\lambda) = 1$]
{\includegraphics[width=0.3\textwidth] {figures/Insta_randomized_subspace_pev_prmanual_degree_glassceiling_avg_final_percplottingxaxis_sns_xlog.pdf}}
\subfloat[Instagram, $f(\lambda) = 1$, multiple EVs]
{\includegraphics[width=0.3\textwidth] {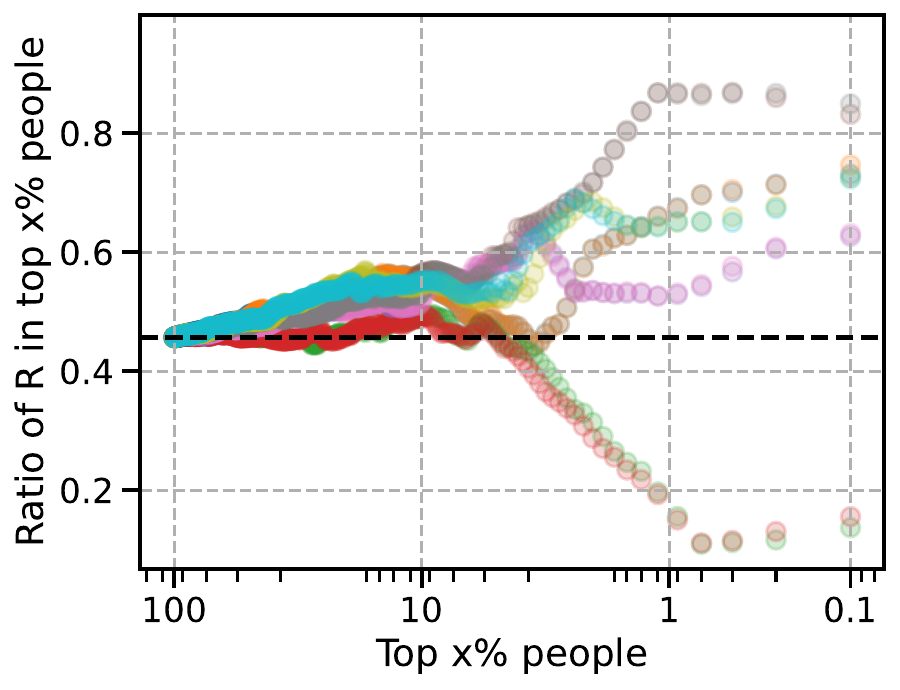}}
\subfloat[Instagram, $f(\lambda) = \lambda^2$, multiple EVs]
{\includegraphics[width=0.3\textwidth] {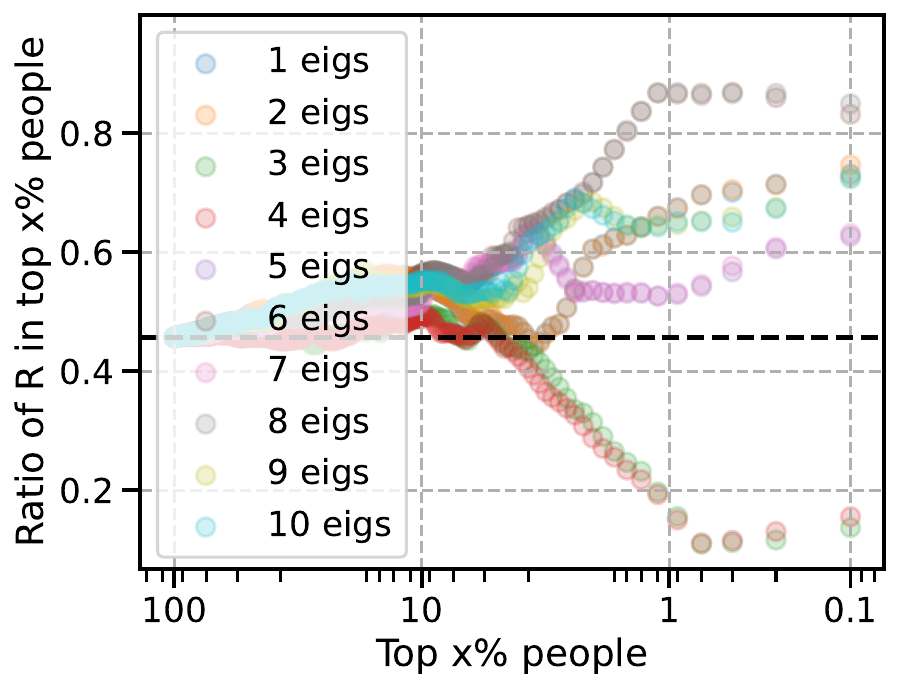}}
\caption{Representation of the minority group R in the ranking of the nodes based on degree (orange), HITS (purple), and Pagerank (blue), for the Instagram dataset, using $6$ eigenvectors and $f(\lambda) = 1$ (a). Figures (b) and (c) show the representation of the minority group when using a varied number of eigenvectors for $f(\lambda) = 1$ (b) and $f(\lambda) = \lambda^2$ (c).}
\label{fig:randomizedhits_ranking_Insta}
\end{figure*}

We experiment with a various number of eigenvectors and aggregations functions $f$ in Figures~\ref{fig:randomizedhits_ranking_APS}--\ref{fig:randomizedhits_ranking_Insta}. Panels (b) and (c) show the percentage of minority present at each rank and above, choosing the first $x$ eigenvectors of the $A^TA$ matrix, as $x$ varies between $1$ and $10$, and aggregating them using $f(\lambda) = 1$ (b) and $f(\lambda) = \lambda^2$ (c). We note that for the larger datasets (DBLP and Instagram), the choice of $f$ between $f(\lambda) = 1$ and $f(\lambda) = \lambda^2$ does not change the ranking, as the gap between different values of the authority scores obtained is larger than the values of the eigenvalues squared. The trends are quite different, depending on the number of eigenvectors chosen, noting that each datasets seems to have a different `optimum' in terms of fairness. For example, for the DBLP dataset, choosing $5$ eigenvectors seems to have the best fairness improvement (and even better than the degree ranking, as Figure~\ref{fig:randomizedhits_ranking_dblpv10} (a) shows); choosing any other number presents no clear pattern.~\citet{cucuringu2011localization} present an empirical analysis of when a minority group gets captured in a lower-order eigenvector, using the \textit{inverse participation ratio} as a measure of a score over eigendirections describing how well-captured a community is in a given eigendirection. They note that a community may appear well-represented in various lower-order eigenvectors. We conclude that using subspace HITS may be in some cases beneficial, but without a stable and consistent pattern for how many eigenvectors might improve fairness for a minority group.~\looseness=-1

\newpage 

\end{document}